\documentclass{aa}
\def\ltsim{\lower.5ex\hbox{$\; \buildrel < \over \sim \;$}}
\usepackage{apj2aa}
\usepackage{astron} 
\usepackage{float,epsfig,psfig}
\usepackage{pstricks}

\begin{document}

\title{XMM-Newton study of the lensing cluster of galaxies CL0024$+$17
\thanks{This work is based on observations
made with the XMM-Newton, an ESA science mission with 
instruments and contributions directly funded by
ESA member states and the USA (NASA).}
\thanks{Based on observations made with the European Southern Observatory
 telescopes obtained from the ESO/ST-ECF Science Archive Facility.}
}
\author{Y.-Y. Zhang\inst{1}, H. B\"ohringer\inst{1},  
Y. Mellier\inst{2}, G. Soucail\inst{3}, and W. Forman\inst{4}
}

\offprints{Y.-Y. Zhang, \\
e-mail: yyzhang@mpe.mpg.de}

\institute{
Max-Planck-Institut f\"ur extraterrestrische Physik,
Giessenbachstra\ss e, 85748 Garching, Germany
\and Institut d'Astrophysique de Paris, 98bis Bd. Arago, 
75014 Paris, France
\and Observatoire Midi-Pyrenees, Laboratorire d'Astrophysique,
MR 5572, 14 Avenue E. Belin, 31400 Toulouse, France
\and Harvard-Smithsonian Center for Astrophysics (CFA), 
60 Garden Street, Cambridge, MA 02138, USA
}

\date{Received 14 May 2004
 / accepted 24 June 2004}
\authorrunning{Zhang et al.}
\titlerunning{XMM-Newton study of the lensing cluster of galaxies CL0024$+$17}

\abstract{We present a detailed gravitational mass measurement
based on the XMM-Newton imaging spectroscopy analysis 
of the lensing cluster of galaxies CL0024$+$17 at $z=0.395$.
The emission appears approximately symmetric. However, 
on the scale of $r \sim 3.3^{\prime}$
some indication of elongation is visible in the northwest-southeast (NW-SE) 
direction from the hardness ratio map (HRM).
Within $3^{\prime}$, we measure a global gas temperature of 
$3.52\pm0.17$~keV, metallicity of 
$0.22\pm0.07$, and bolometric luminosity 
of $2.9 \pm 0.1 \times 10^{44} h^{-2}_{70}$~erg~s$^{-1}$.
We derive a temperature distribution with an 
isothermal temperature of 3.9~keV to a radius of $1.5^{\prime}$
and a temperature gradient in the outskirts 
($1.3^{\prime}<r<3^{\prime}$).
Under the assumption of hydrostatic equilibrium, we 
measure gravitational mass and gas mass fraction to be
$M_{200}=2.0 \pm 0.3 \times 10^{14}$~$h_{70}^{-1}{\rm M_{\odot}}$ 
and $f_{\rm gas}=0.20\pm0.03 h^{-3/2}_{70}$ at 
$r_{200}=1.05 h^{-1}_{70}$~Mpc
using the observed temperature profile. 
The complex structure in the core region is the key to explaining  
the discrepancy in gravitational mass 
determined from XMM-Newton X-ray observations 
and HST optical lensing measurements.

 }

\maketitle

\section{Introduction}

\begin{figure}
\begin{center}
\includegraphics[width=8cm]{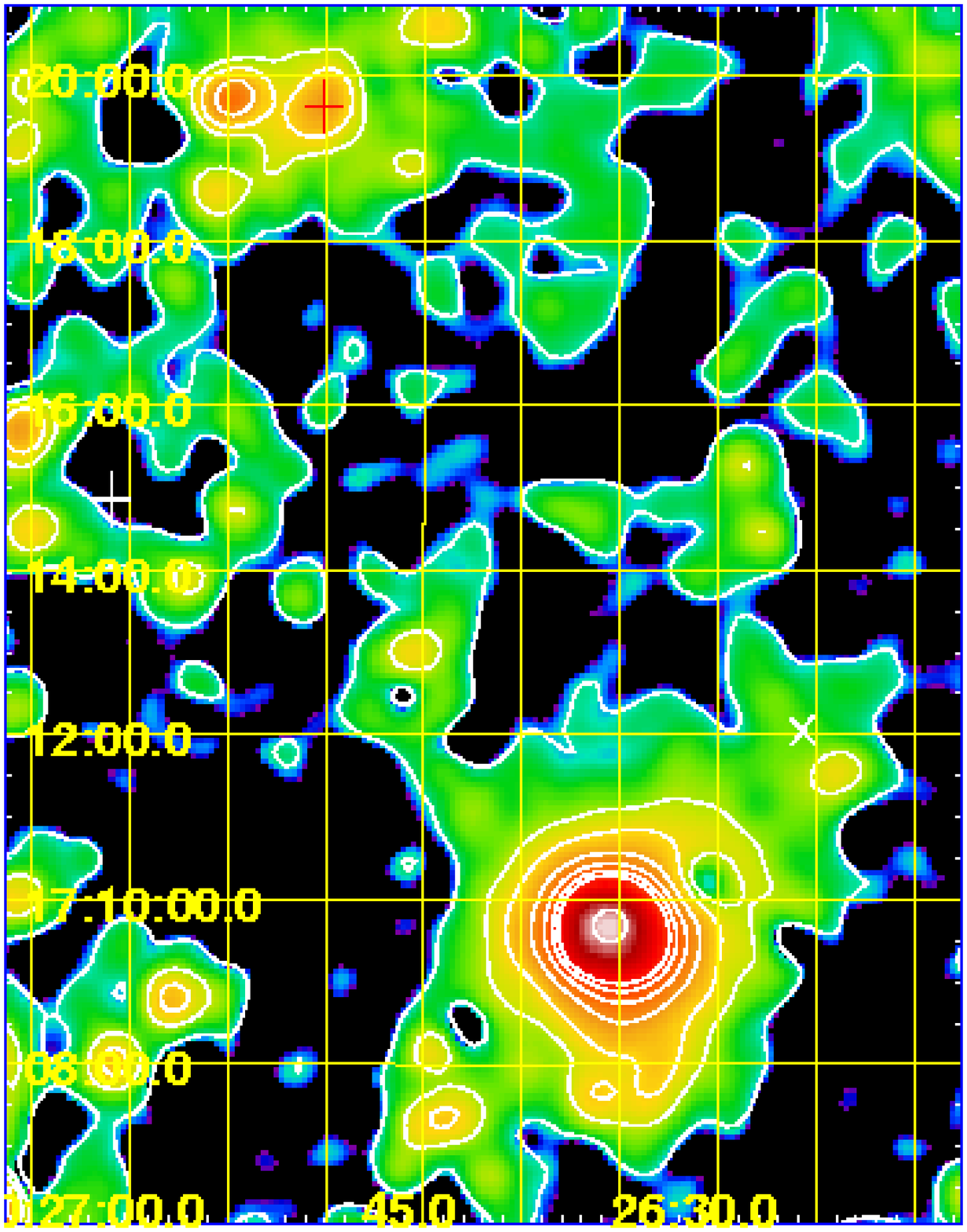}
\end{center}
\figcaption{Flat fielded, point sources masked and 
adaptively-smoothed XMM-Newton 
image of CL0024$+$17 in the 0.5--2~keV 
band with logarithmically scaled contours. 
Additionally we show the galaxy pair (cross) 
in Czoske et al. (2001; 2002) and the substructure (X-point) 
in kneib et al. (2003). 
\label{f:image}
}
\end{figure}
\begin{figure*}
\begin{center}
\includegraphics[width=8.5cm]{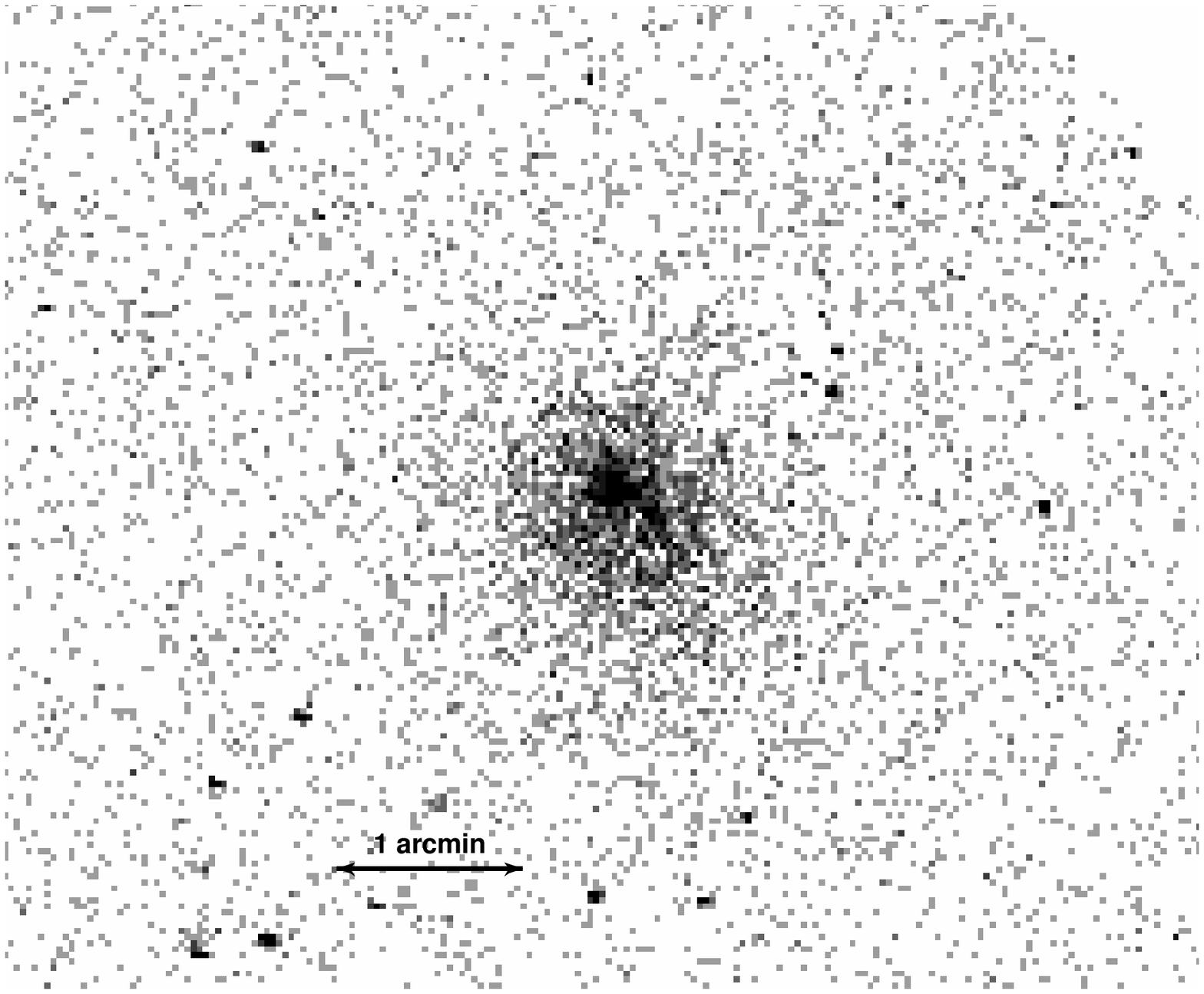}
\includegraphics[width=8.5cm]{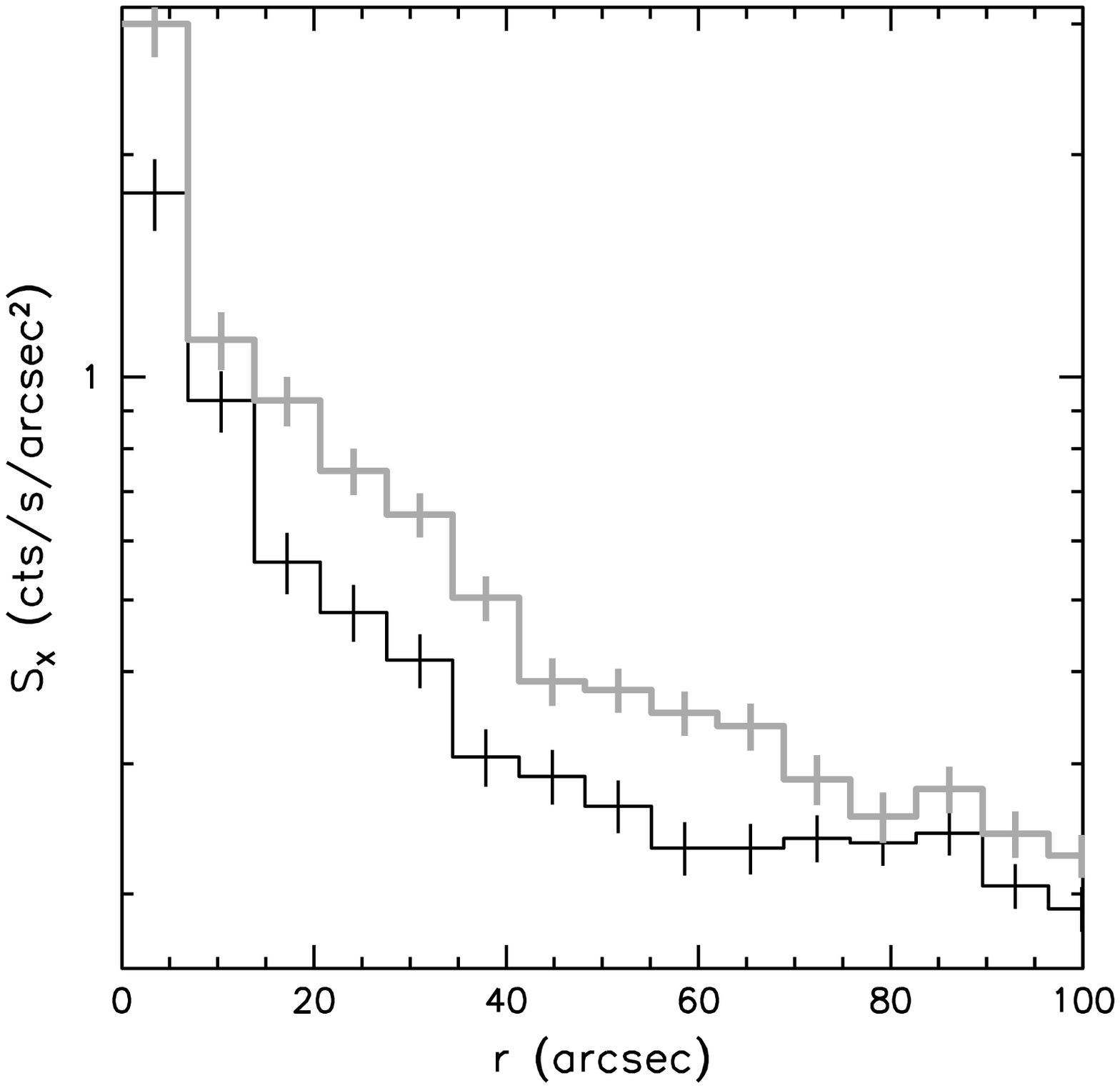}\\
\end{center}
\figcaption{
Chandra image in the 0.5--2.5~keV band (left) and 
surface brightness profiles (right) in annuli using 
the 0.5--2.5~keV image with azimuths 30--120$^{\circ}$ 
(counter clockwise) from north to east (black) 
and from south to west (grey), respectively.
\label{f:cdf}
}
\end{figure*}
\begin{figure*}
\begin{center}
\includegraphics[width=17.cm]{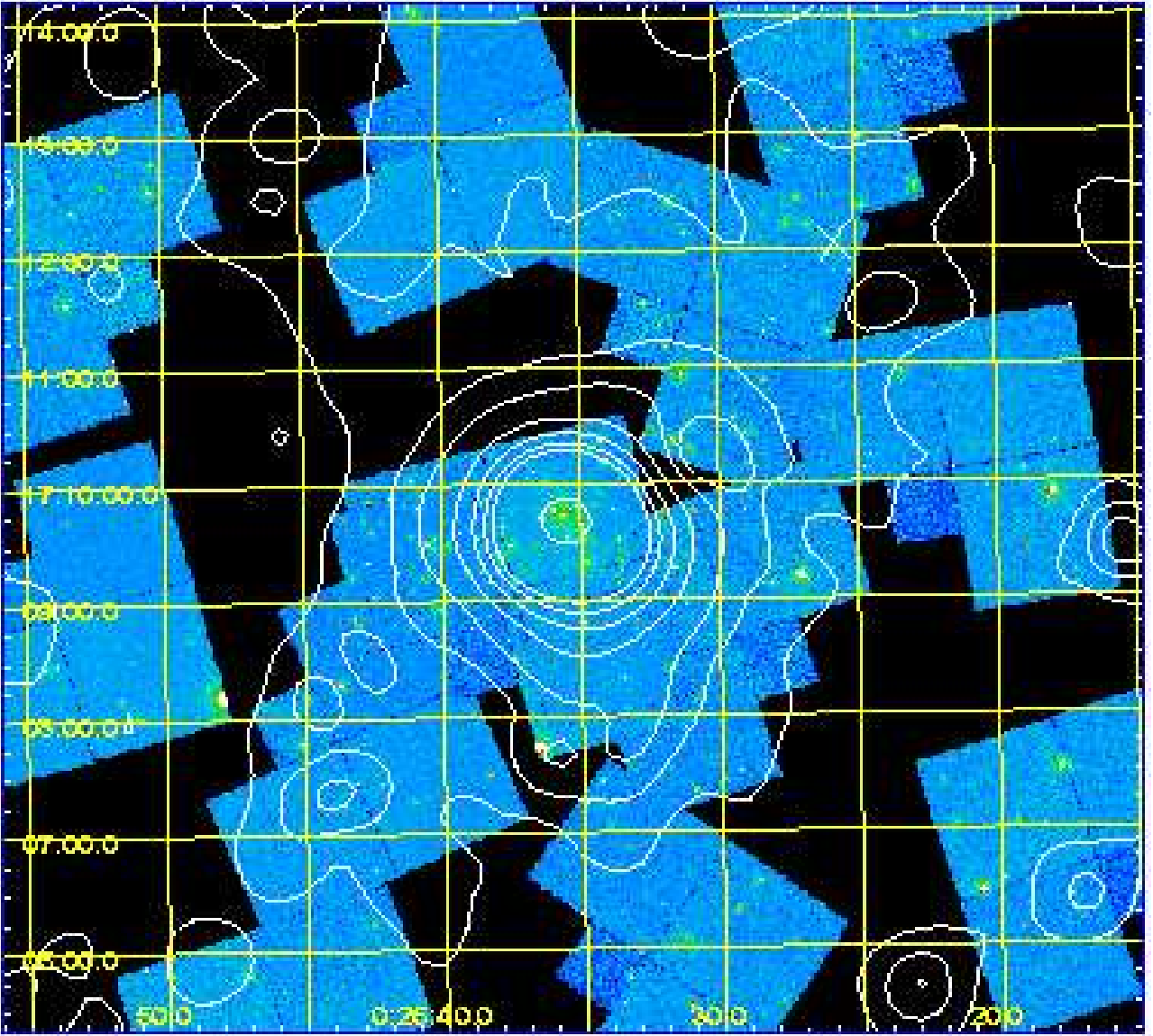}
\end{center}
\figcaption{HST mosaic image
of the X-ray center of CL0024$+$17.  
Superposed X-ray contours of XMM-Newton
as shown in Fig.~\ref{f:image}
indicate the elongation in 
the NW-SE direction on the scale of $r \sim 3.3^{\prime}$ (X-point). 
\label{f:hstimage}
}
\end{figure*}  
\begin{figure}
\begin{center}
\includegraphics[width=8.5cm]{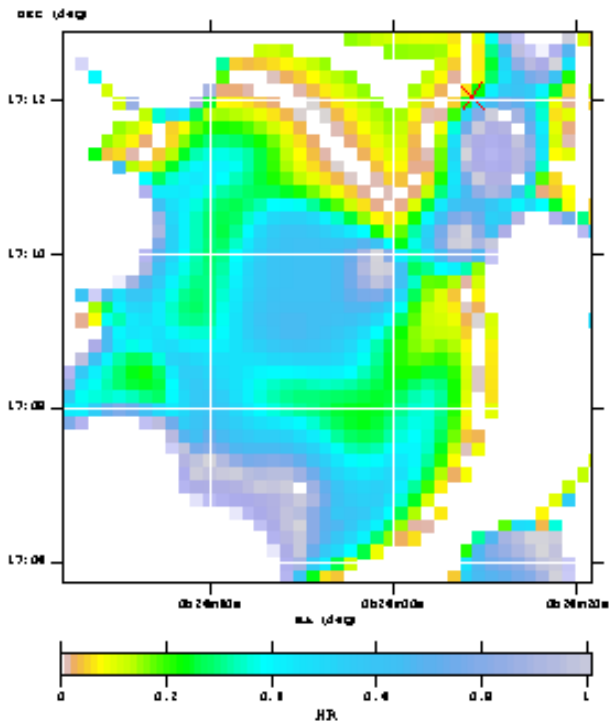}
\end{center}
\figcaption{XMM-Newton HRM ($\sim$10$^{\prime \prime}$ pixels) of CL0024$+$17.
Additionally we show the position of the substructure (X-point) 
in kneib et al. (2003).  
\label{f:hnr}
}
\end{figure}  
\begin{figure}
\begin{center}
\includegraphics[width=8.5cm,angle=0]{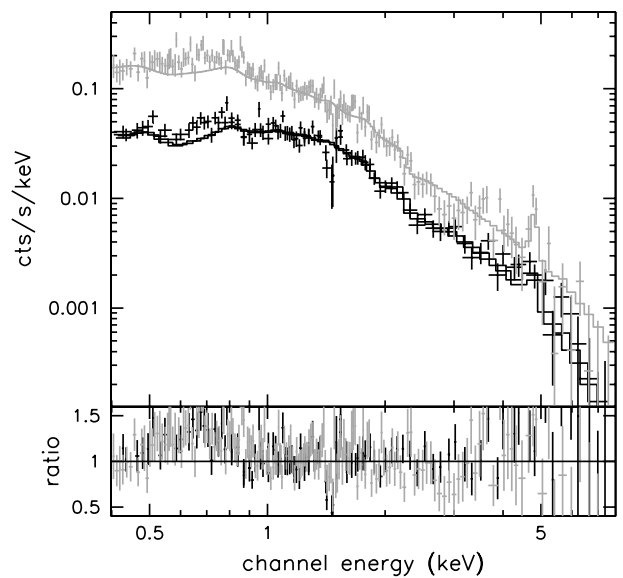}
\end{center}
\figcaption{XMM-Newton spectra (top panel) extracted 
from the $<3^{\prime}$ region fitted in the 1--10~keV band 
by an isothermal model 
and their residuals (bottom panel)
for MOS (black) and pn (grey), respectively.
\label{f:tempha}}
\end{figure}
\begin{figure*}
\begin{center}
\includegraphics[width=8.5cm]{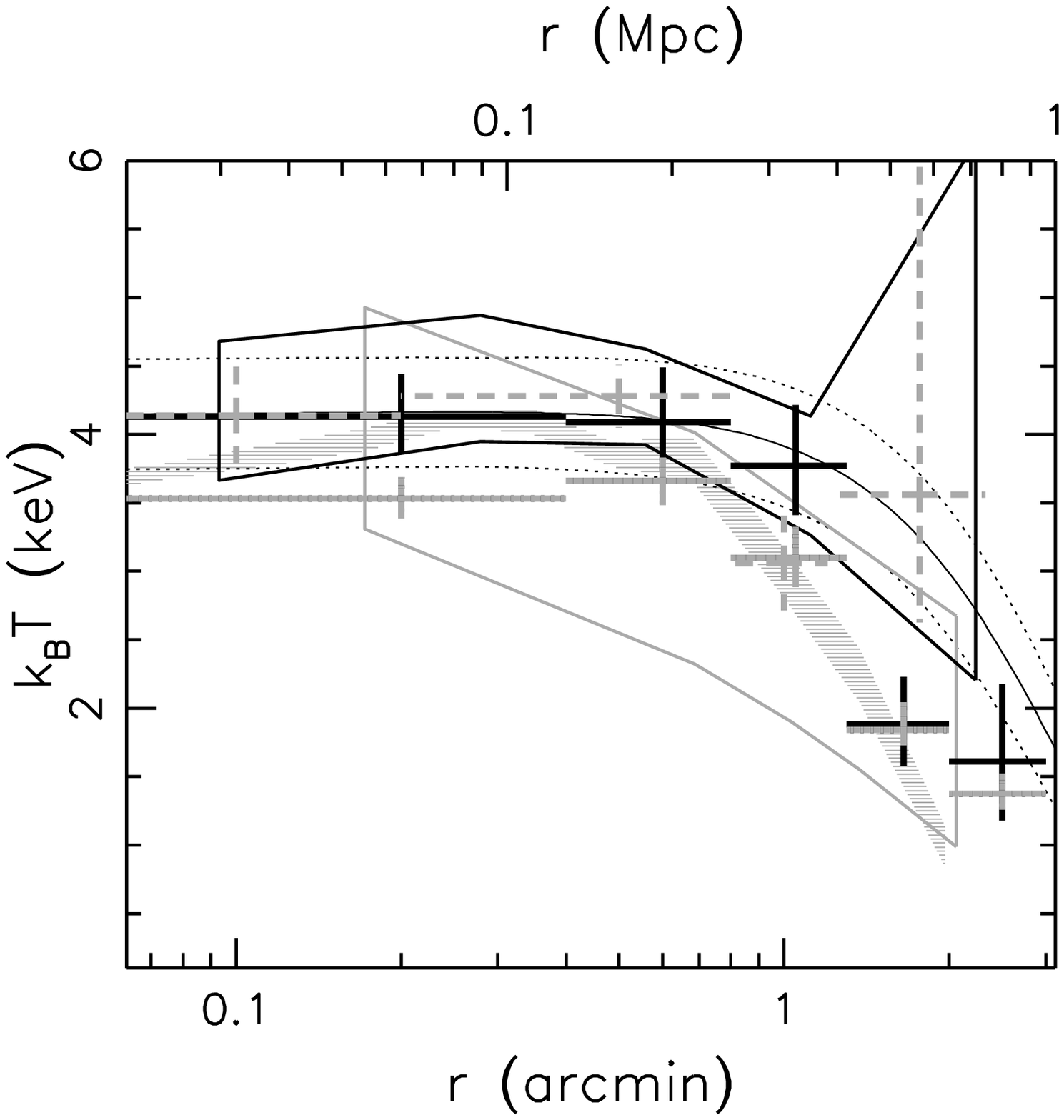}
\includegraphics[width=8.5cm]{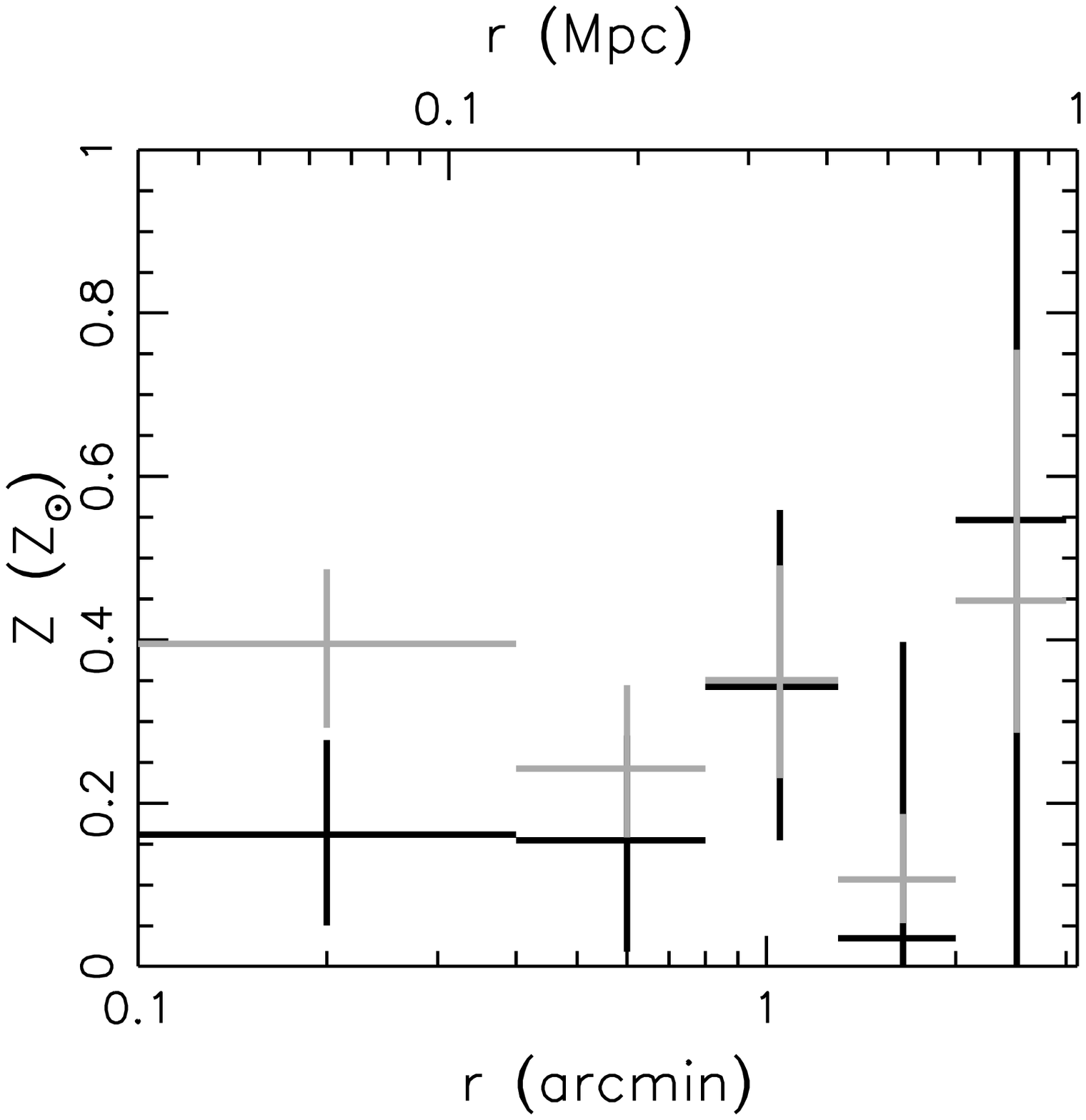}\\
\end{center}
\figcaption{Projected 
temperature (left) and metallicity (right) profiles 
measured from the XMM-Newton data in the 1--7.8~keV band (solid, black) 
and 0.4--7.8~keV band (solid, grey) with
point sources excluded.
A solid curve in black indicates
the best Gaussian fit of the 
temperature measurements in the 1--7.8~keV band 
with confidence intervals as dotted curves. 
Additionally the similarity temperature profiles in Markevitch 
et al. (1998; grey) and in Zhang et al. (2004b; black) are 
indicated by solid outlines.  
Our Chandra temperature measurements of CL0024$+$17 (dashed, grey) and 
the temperature profile of Sersic~159$-$03 (grey shadow; 1 $\sigma$ interval) 
in Kaastra et al. (2001) scaled to CL0024$+$17 are presented for comparison.
\label{f:ktZ}
}
\end{figure*}
\begin{figure}
\begin{center}
\includegraphics[width=7.5cm,,angle=270,clip]
{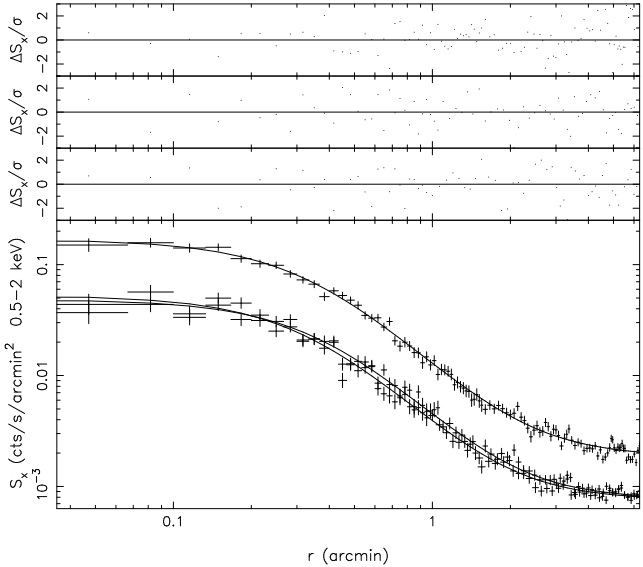}
\end{center}
\figcaption{Flat fielded, 
point source subtracted and
azimuthally averaged radial surface brightness profiles 
for CL0024$+$17 in the 0.5--2~keV band 
for pn (top) and MOS (bottom) and their best $\chi^2$ fits
by a PSF convolved $\beta$-model combining a constant  
residual soft X-ray background. 
Residuals scaled by the data uncertainties 
appear in the upper three panels for MOS1, MOS2 and pn, 
respectively, from top to bottom. 
\label{f:sx}
}
\end{figure}
\begin{figure}
\begin{center}
\includegraphics[bb=100 375 540 687,width=8.5cm,clip]
{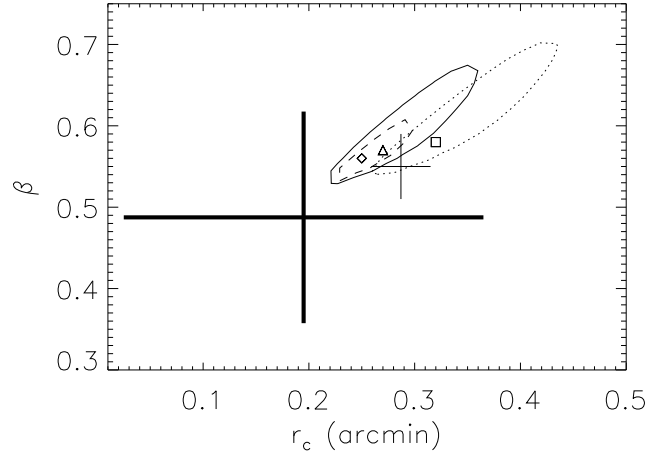}
\end{center}
\figcaption{
90\% confidence contours and the best fit values for the core 
radius $r_{\rm c}$ and slope parameter $\beta$ of the $\beta$ model for 
the surface brightness of XMM-Newton for MOS1 (solid; triangle), 
MOS2 (dotted; square) and pn (dashed; diamond), respectively. 
Chandra (thin; Ota et al. 2004) and ROSAT (thick; B\"ohringer et al. 2000) 
measurements are also shown as crosses.
\label{f:contour}
}
\end{figure}
\begin{figure}
\begin{center}
\includegraphics[width=8.5cm]
{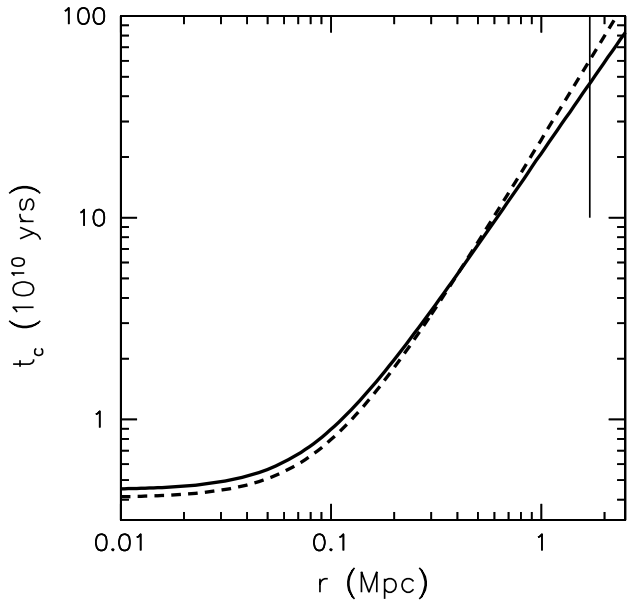}
\end{center}
\figcaption{Cooling time of the gas using
the global temperature (dashed) and
using the observed temperature profile (solid).
The vertical line indicates $r_{200}=1.7 h^{-1}_{70}$~Mpc,
predicted from the lensing data in Kneib et al.(2003).
\label{f:tc}
}
\end{figure}
\begin{figure*}
\begin{center}
\includegraphics[width=8cm]{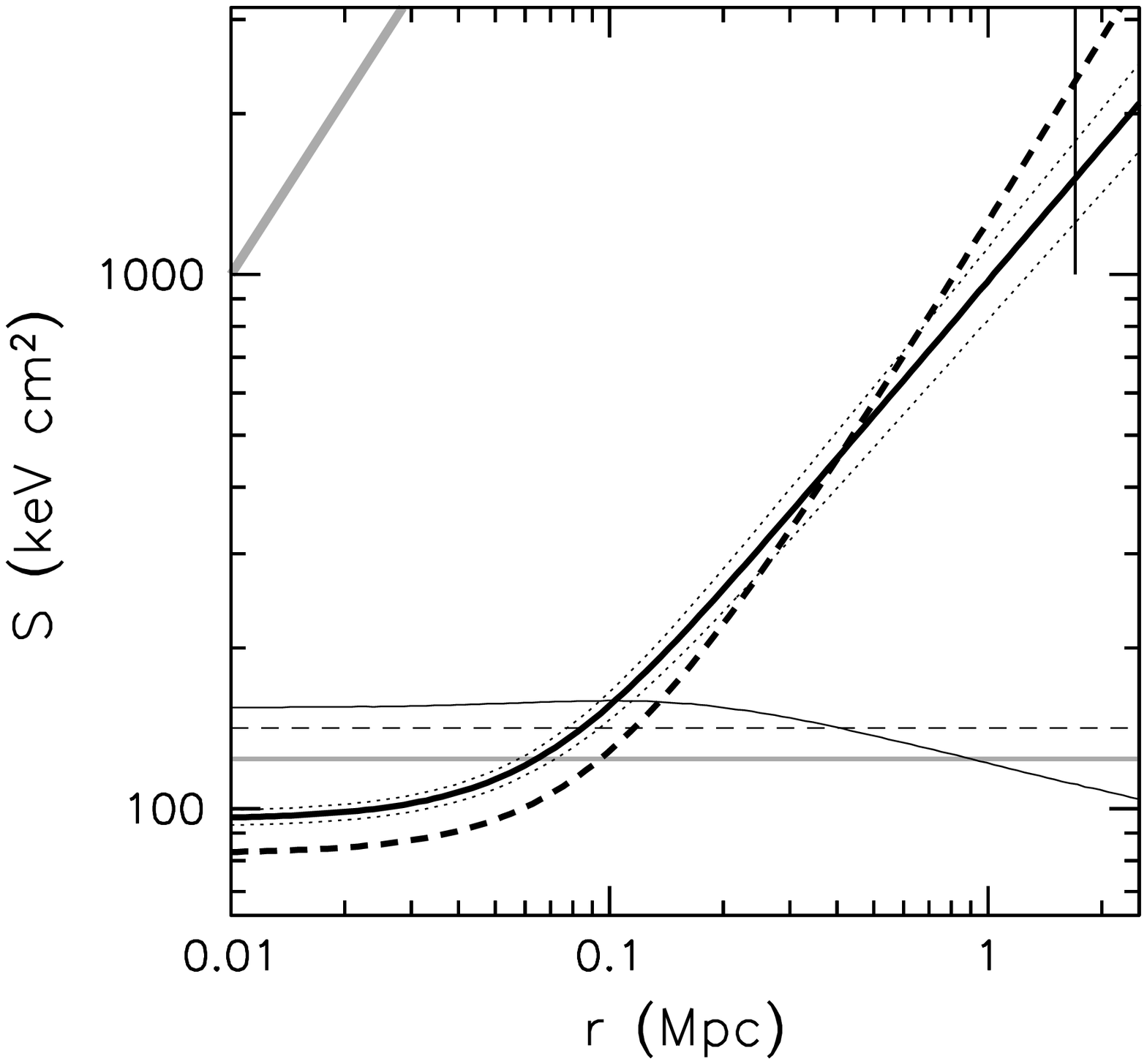}
\includegraphics[width=8cm]{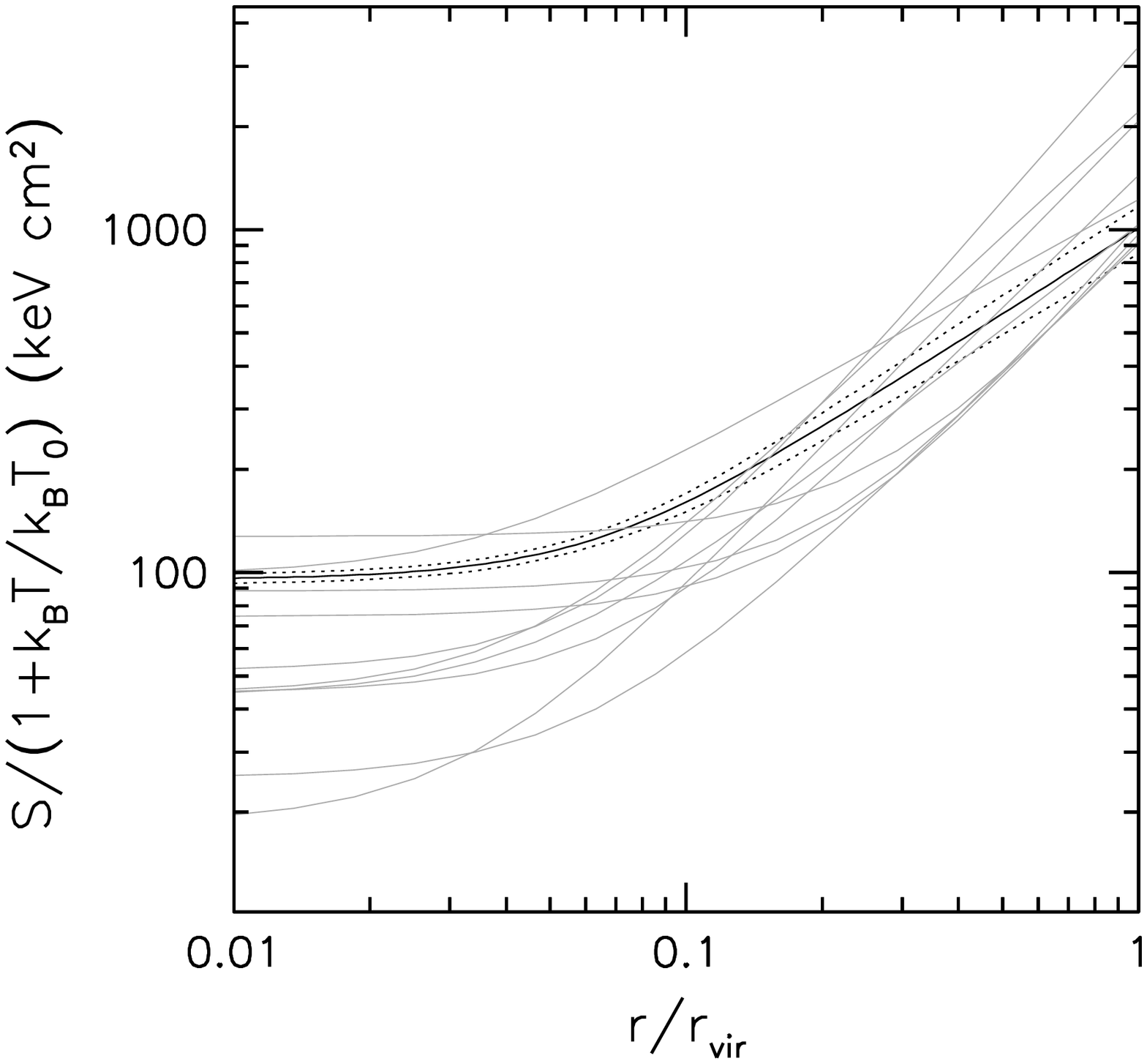}
\end{center}
\figcaption{Left: 
Entropy (thick, black) and critical entropy floor (thin, black) 
using the global temperature (dashed) and
using the observed temperature profile
(solid, confidence intervals as dotted curves).
Additional grey lines 
are the entropy floor (thin), $S\sim 124 h^{-1/3}_{70}$~${\rm keV~cm^2}$, 
from Lloyd-Davies et al. (2000)
and the predicted slope of 1.1 (thick) from the 
spherical accretion shock model (Kay 2004).
The vertical line indicates the radius as defined in Fig.~\ref{f:tc}.
Right: Scaled entropy of CL0024$+$17 vs. scaled radius
using the radius of $r_{200}=1.05 h^{-1}_{70}$~Mpc
(black, confidence intervals as dotted curves) and 
the Birmingham-CfA clusters in a temperature
range of 2.9--4.6~keV (grey).
\label{f:scaleds}
}
\end{figure*}
\begin{figure*}
\begin{center}
\includegraphics[width=6.8cm]{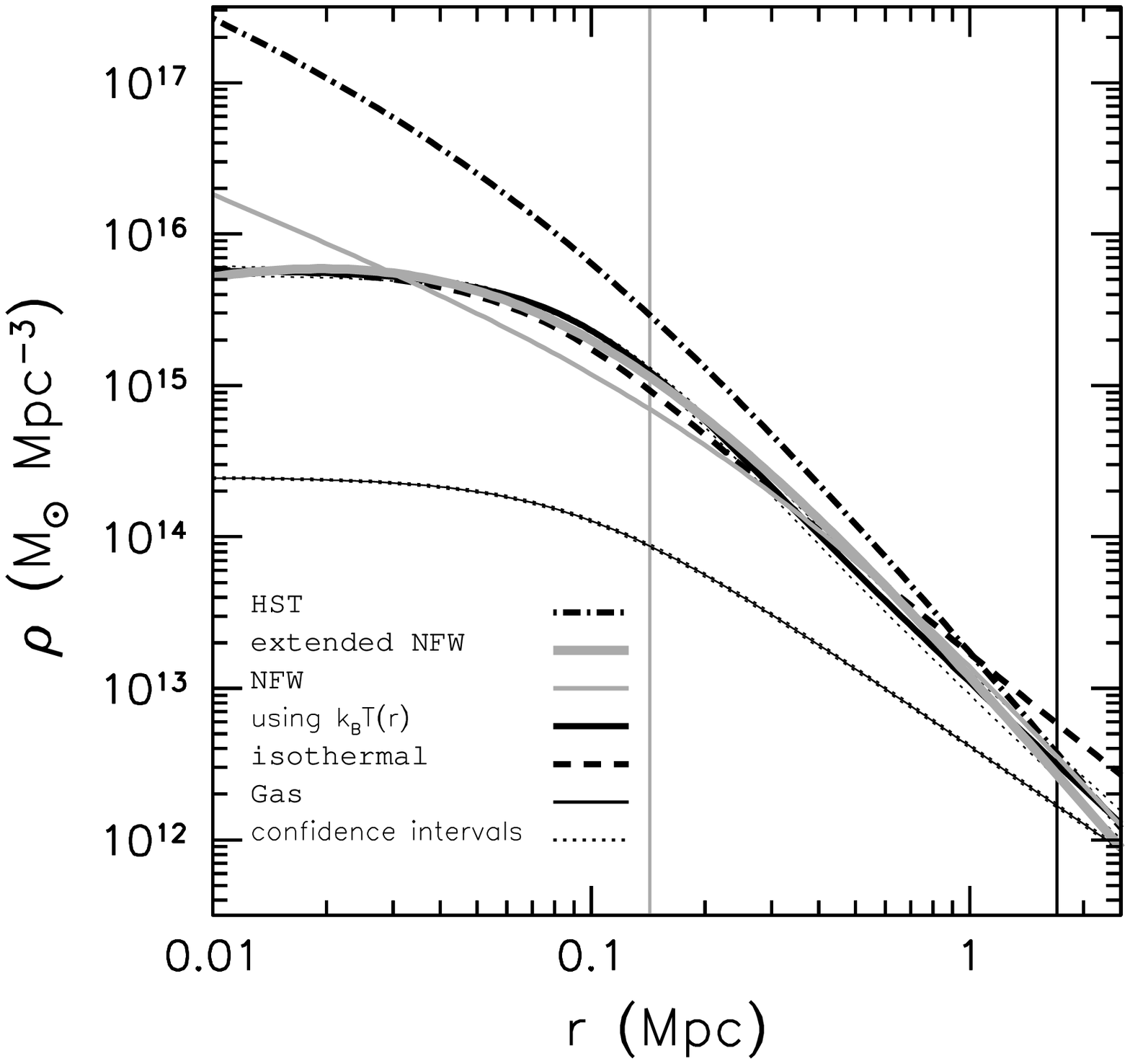}
\includegraphics[width=6.8cm]{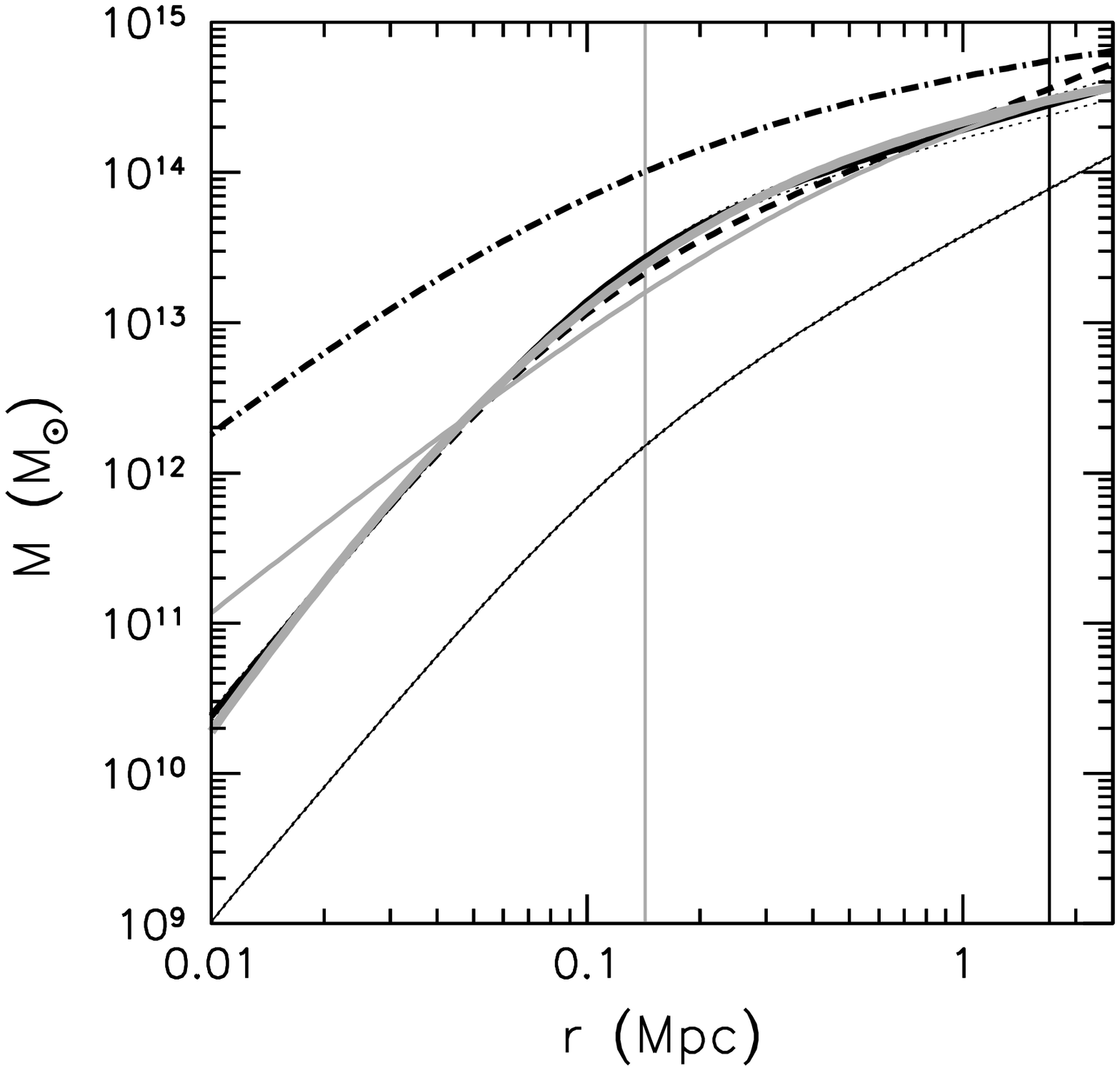}\\
\includegraphics[width=6.8cm]{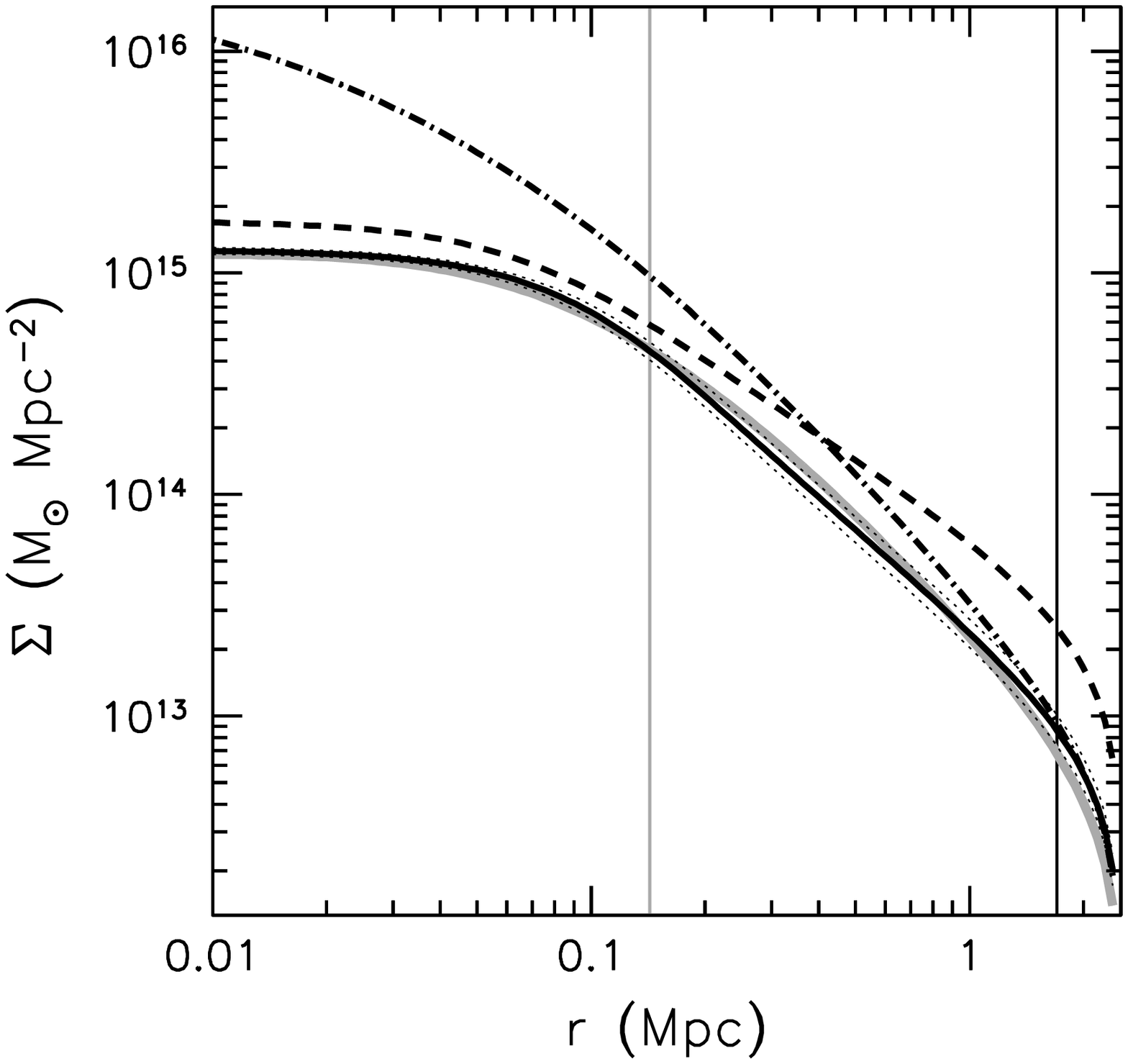}
\includegraphics[width=6.8cm]{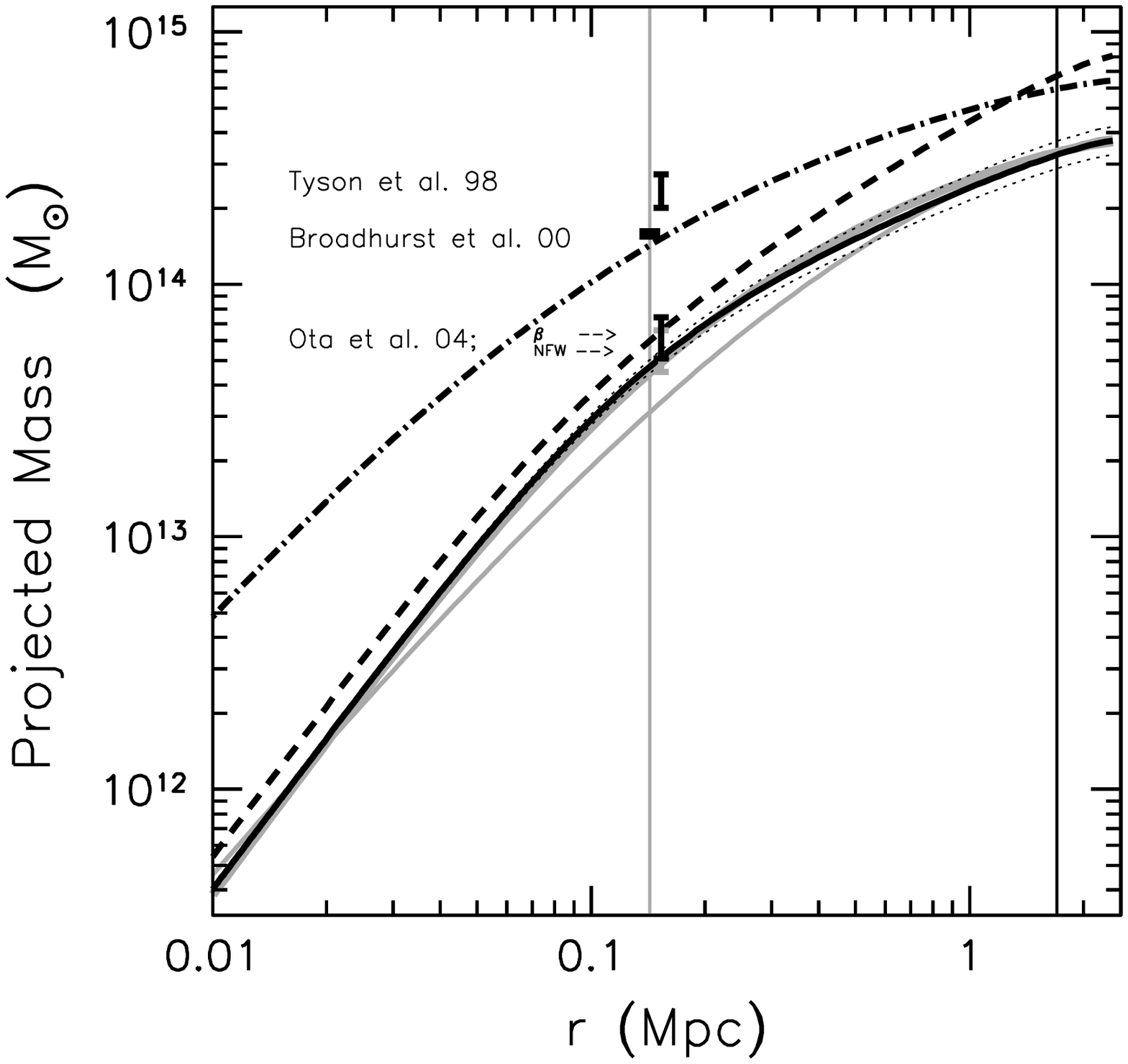}\\
\includegraphics[width=6.8cm]{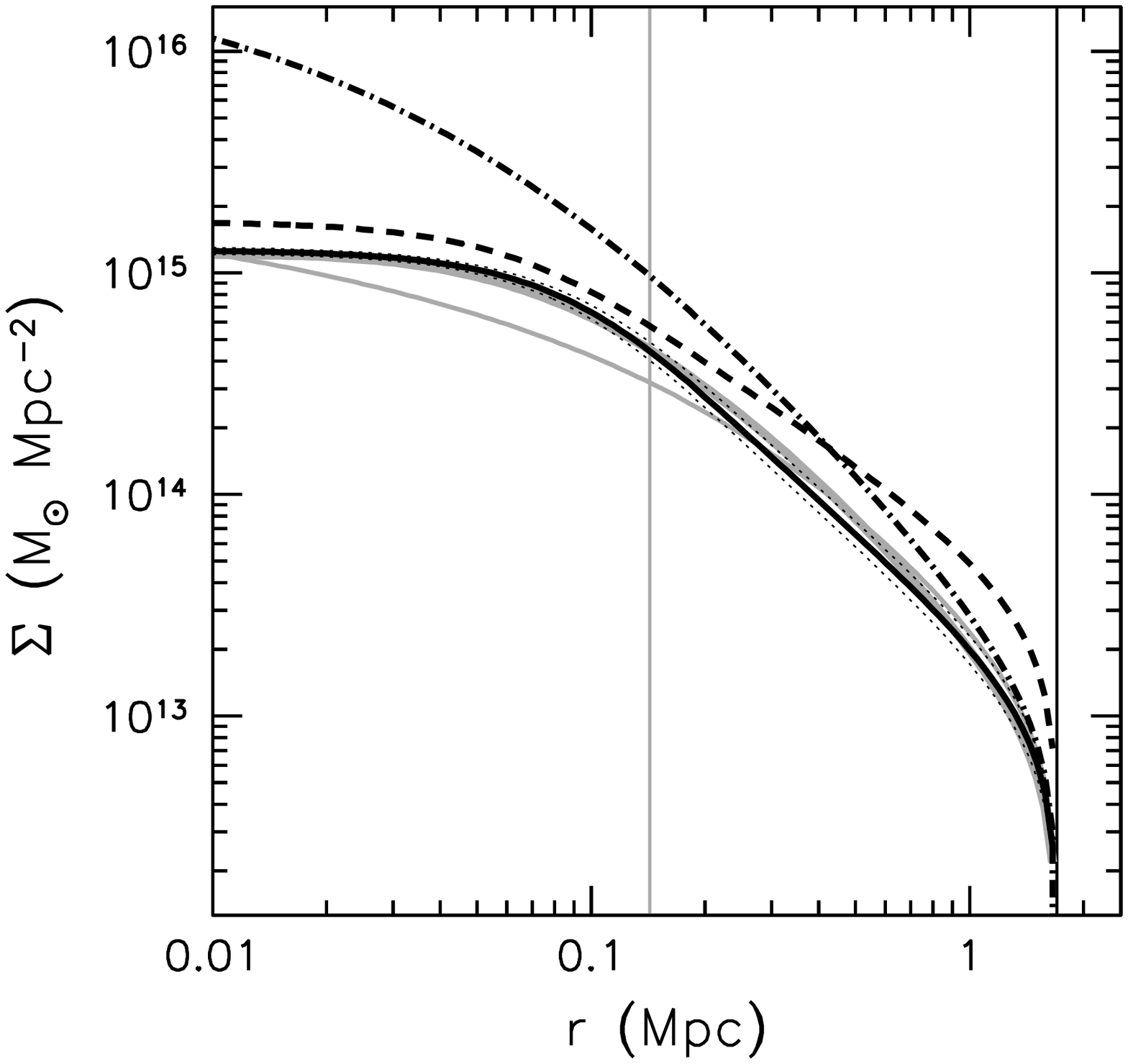}
\includegraphics[width=6.8cm]{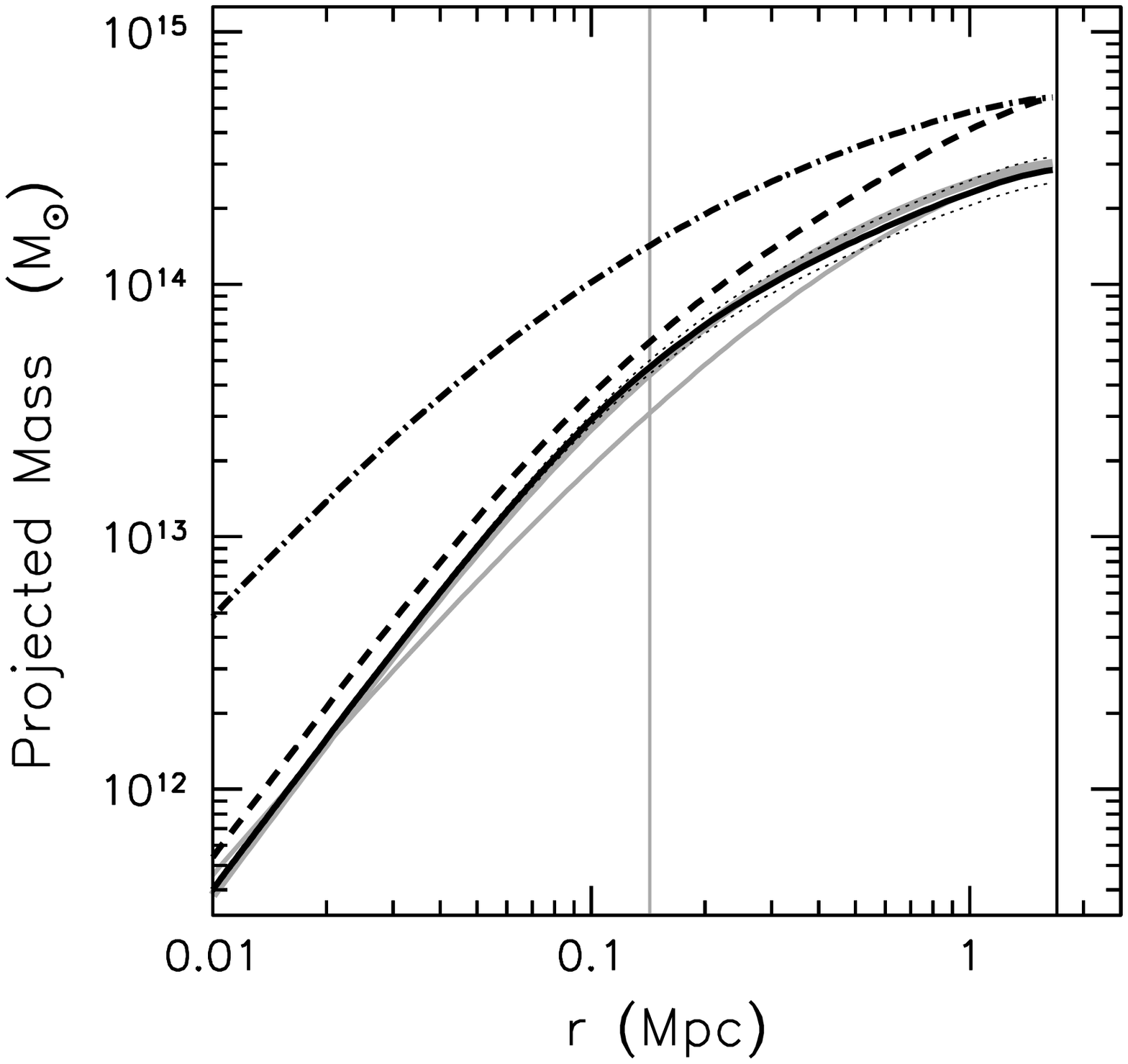}
\end{center}
\figcaption{
{\bf Upper left:} dark matter density and gas density, 
{\bf upper right:} gravitational mass and gas mass, 
{\bf middle left:} projected dark matter density 
($r_{\rm truncate}=2.5 h^{-1}_{70}$~Mpc), 
{\bf middle right:}  projected gravitational mass
($r_{\rm truncate}=2.5 h^{-1}_{70}$~Mpc), 
{\bf lower left:} projected dark matter density 
($r_{\rm truncate}=1.7 h^{-1}_{70}$~Mpc), 
{\bf lower right:}  projected gravitational mass 
($r_{\rm truncate}=1.7 h^{-1}_{70}$~Mpc).
The meaning of the lines is: (i) hydrostatic equilibrium (black) using
the global temperature  
(thick, dashed) and 
using the observed temperature profile
(thick, solid, confidence intervals as dotted curves) and its
best fit (grey) of extended NFW model (thick) and NFW model 
(thin);
(ii) HST lensing measurements 
(Kneib et al. 2003; dash-dotted).
Additionally we present the gas density and gas mass
(thin, black) with confidence intervals as dotted curves.
We label the strong lensing projected gravitational mass  
of $1.59 \pm 0.04 \times 10^{14} h^{-1}_{70}{\rm M_{\odot}}$ 
(Broadhurst et al. 2000) and 
$ 2.37 \pm 0.36 \times 10^{14} h^{-1}_{70}{\rm M_{\odot}}$
(Tyson et al. 1998) at the arc radii of 
$0.143 h^{-1}_{70}$~Mpc and $0.153 h^{-1}_{70}$~Mpc 
and the Chandra results (Ota et al. 2004) using a $\beta$/NFW model 
at the arc radius of $0.153 h^{-1}_{70}$~Mpc for comparison. 
The black vertical line indicates the radius as defined 
in Fig.~\ref{f:tc}.
The grey vertical line indicates the arc radius of 
$0.143 h^{-1}_{70}$~Mpc (e.g. Broadhurst et al. 2000).
\label{f:mass}
}
\end{figure*}
\begin{figure}
\begin{center}
\includegraphics[width=8.5cm]
{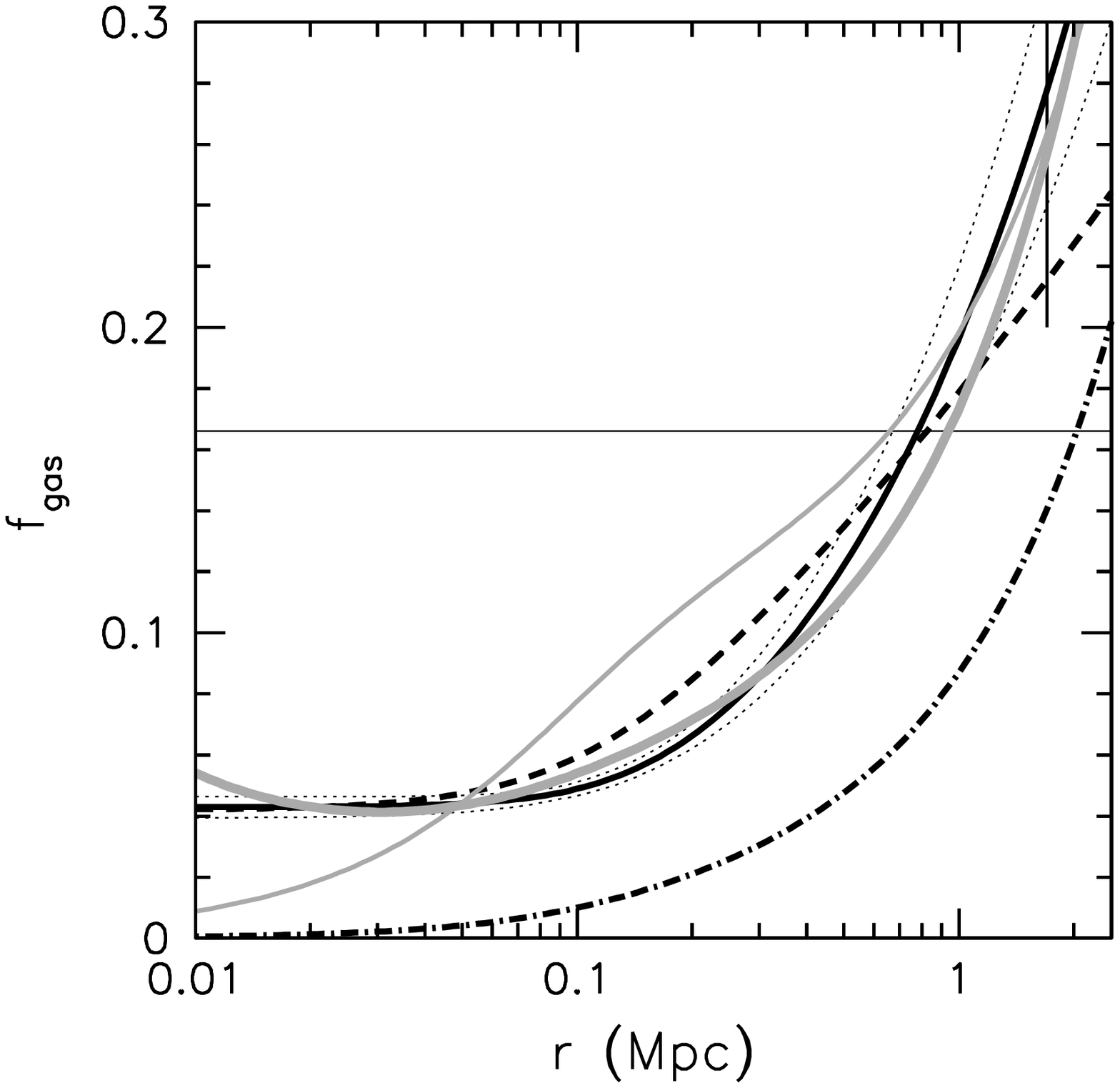}
\end{center}
\figcaption{Gas mass fraction as a function 
of radius. The lines have an identical meaning to those in Fig.~\ref{f:mass}
except that
an additional horizontal line indicates
the WMAP measurement $f_{\rm b}=\Omega_{\rm b}/\Omega_{\rm
m}=0.166$ (Spergel et al. 2003).
\label{f:fg}
}
\end{figure}

One of the optically most prominent, but also most puzzling distant
lensing galaxy clusters, is CL0024$+$17 at a redshift of $z=0.395$
(Gunn \& Oke 1975).
Its high galaxy density led to the early discovery by 
Humason \& Sandage (1957), and the cluster has since been the subject of many 
studies. It was one of the first clusters to display 
the so-called Butcher-Oemler effect (Butcher \& Oemler 1978;
Dressler \& Gunn 1982; Dressler et al. 1985), and 
Schneider et al. (1986) described it as a very rich optical cluster.

Koo (1988) discovered gravitational arcs in CL0024$+$17,
and it was subsequently studied extensively
(Mellier et al. 1991; Kassiola et al. 1992; Kassiola et al. 1994;
Wallington et al. 1995; Colley et al. 1996;
Smail et al. 1997; Tyson et al. 1998; Broadhurst et al. 2000;
Shapiro \& Iliev 2000; Treu et al. 2003; Kneib et al. 2003).  
A total of eight arc-like lensed images from the same background 
galaxy were identified (e.g. Tyson et al. 1998).
The redshift of the lensed galaxy was determined by Broadhurst et al.
(2000) to be $z=1.675$. 
Bonnet et al. (1994) detected a weak shear signal in this cluster 
out to a radius of $2.1~h^{-1}_{70}$~Mpc
and inferred a high gravitational mass of about 
1.4--2.9$\times 10^{15}~h^{-1}_{70}{\rm M_{\odot}}$.
Recently Kneib et al. (2003) revised this lensing mass estimate
downward on the basis of a very detailed Hubble Space Telescope 
(HST) mapping of the cluster which yielded a gravitational mass of
$M_{200}=5.7^{+1.1}_{-1.0} \times 10^{14} h^{-1}_{70}{\rm M_{\odot}}$
out to $r_{200}=1.7 h^{-1}_{70}$~Mpc.
In spite of the massive appearance of CL0024$+$17
at optical wavelengths, the cluster is relatively 
faint in X-rays with a ROSAT determined luminosity of 
$L_{\rm X}=1.22\pm 0.08\times 10^{44} h^{-2}_{70}$~erg~s$^{-1}$
in the 0.1--2.4~keV band (B\"ohringer et al. 2000),
an ASCA determined temperature of $5.7^{+4.2}_{-2.1}$~keV 
(Soucail et al. 2000),
and a Chandra temperature of 
$4.47^{+0.42}_{-0.27}$~keV (Ota et al. 2004;
$M_{200}=4.6^{+0.7}_{-0.5} \times 10^{14} h^{-1}_{70}{\rm M_{\odot}}$
to $r_{200}=1.4 h^{-1}_{70}$~Mpc).
In addition to the surprisingly low X-ray luminosity in comparison 
to its optical prominence, 
the detailed analysis of the X-ray observations yield a total 
cluster mass in the range 
$M_{200}=2$--$4.6 \times 10^{14}$~$h_{70}^{-1}{\rm M_{\odot}}$ 
that is 1.3--3 times lower than the weak lensing determined
mass (Soucail et al. 2000; B\"ohringer et al. 2000; Ota et al. 2004).  
The discrepancy is even larger at the arc radius 
$\sim 0.143 h^{-1}_{70}$~Mpc (e.g. Broadhurst et al. 2000).

A hint of the explanation for 
this mass discrepancy came from the detailed analysis 
of the galaxy dynamics in CL0024$+$17 based on $\sim 300$ redshifts 
of galaxy members (Czoske et al. 2001; 2002).
Czoske et al. (2001; 2002) found that the line-of-sight velocity distribution 
is not that of a relaxed cluster and is at least bimodal. They also
demonstrated that the redshifts can approximately be explained 
by a line-of-sight merger of two systems with a mass ratio of 
the order of 1:2.
Also the weak lensing analysis shows a mass distribution
with substructure, modeled as a bimodal distribution of
two systems with slightly different central positions in the plane of the sky 
(Kneib et al. 2003).  
The infrared observations 
imply significant star forming activity
with Star Formation Rates (SFRs) one to two orders of magnitude
higher than those computed from the optical.
The underestimation of the SFRs in the optical is due to 
absorption by dust (Coia et al. 2003).

Using the capability of XMM-Newton to perform imaging 
spectroscopy at high angular resolution and thus allowing us to study 
the density and temperature structure of the intracluster
medium (ICM), we performed an XMM-Newton observation of 
CL0024$+$17 to shed new light on this enigmatic system. 

This paper is structured as follows.  In Sect.~\ref{s:method}, we
describe a double background subtraction method, which
is developed to provide precise spectral background removal. In
Sect.~\ref{s:result}, we analyse the properties of the hot gas in the
galaxy cluster CL0024$+$17, then
determine the total mass, projected gravitational mass, 
and gas mass fraction based on precise temperature and gas density profiles. 
In Sect.~\ref{s:discussion}, we discuss the structures in the cluster 
and possible solutions for the discrepancy between the X-ray and HST 
lensing measurements in the gravitational mass.
In Sect.~\ref{s:conclusion}, we present our conclusions.  We
adopt a flat $\Lambda$CDM cosmology with density parameter
$\Omega_{\rm m}=0.3$ and Hubble constant $H_{\rm
0}=70$~km~s$^{-1}$~Mpc$^{-1}$. Thus 
$1^{\prime}=0.320 h^{-1}_{70}$~Mpc. All coordinates are given in epoch J2000.
We adopt the solar abundance values of Anders \& Grevesse (1989).
Error bars correspond to the 68\%
confidence level, unless explicitly stated otherwise.

\section{Method}
\label{s:method}

\subsection{Data preparation}

CL0024$+$17 was observed on Jan 6th, 2001 for a total exposure 
of 52.1~ks, 52.1~ks and 48.3~ks for MOS1, MOS2 and pn, 
respectively, by XMM-Newton 
with the European Photon Imaging Camera (EPIC) 
in standard Full Frame (FF) mode and Extended Full Frame (EFF)
mode for MOS and pn, respectively. 
For all detectors, the thin filter was used.
Data reduction and calibration was carried out using 
the XMM-Newton Science Analysis System (SAS5.4.1).
The central position of the observation
is ${\rm R.A.}=00^h26^m35^s.7$, ${\rm decl.}=17^{\circ}09^{\prime}35^{\prime\prime}.8$. 

Above 10~keV, there is little X-ray emission from the cluster due to
the low telescope efficiency. The particle
background therefore completely dominates.  The cluster emission is
not variable, so any spectral range can be used for temporal variability
studies of the background.  
The 10--12~keV (12--14~keV) energy band (binned in 100~s
intervals) was used to monitor the particle background and to excise
periods of high particle flux for MOS (pn). 
In this screening process we use the
settings $FLAG=0$ and
$PATTERN \le 12$ ($PATTERN \le 4$) for MOS (pn).

We reject those time intervals affected by flares 
in which the detector countrate (ctr)
exceeds a threshold of $2 \sigma$ above the average ctr,
where the average and the variance have been iteratively determined 
from the ctr histogram below the rejection threshold.
We screened the data using the thresholds of 21.4, 22.3 and 56.2 for 
MOS1, MOS2 and pn. The net exposure time is 46.6~ks, 46.0~ks and 
42.9~ks for MOS1, MOS2 and pn, respectively.

\subsection{Background analysis}
\label{s:mos}

We do not expect any cluster emission outside $5.4^{\prime}$, the 
virial radius estimated from the redshift (z=0.395; Gunn \& Oke 1975) and 
the Chandra average temperature measurement
(4.47~keV, Ota et al. 2004). Thus the cluster 
emission covers less than half of the field
of view (FOV) of the XMM-Newton telescope detectors.
Actually the significant cluster emission extends only to $3^{\prime}$ 
in the surface brightness of the XMM-Newton observations.   
Thus we use the $5<r<5.5^{\prime}$ region in detector coordinates to 
compare the background conditions of
the XMM-Newton pointing of the Chandra Deep Field South 
(CDFS, which we use as background field for the spectral analysis) 
with that of CL0024$+$17. 
The background of CL0024$+$17 is higher than that of 
the background field (CDFS) by 
$\sim 30\%$ for MOS and $\sim 10\%$ for pn on average in the 
complete energy band, in which the difference is mainly produced   
in the low energy band below 1~keV. 

In the spectral analysis, the regions not affected by cluster 
emission, enable us to study the residual background in the 
source observation compared to the background observation.
We applied a double background subtraction method as 
described in Zhang et al. (2004a) using the results of the 
residual background modeling.
In summary the spectral analysis is performed in two steps
using the software package XSPEC11.3.0:
(i)  A power law model for the residual background (background
difference) is obtained in XSPEC from a comparison
of the outer region (5--5.5$^{\prime}$) of the target 
and background fields (cf. Table~\ref{t:powb}).
(ii) The spectral modeling is performed in XSPEC with the cluster 
region of the target field as source data, the region 
in the same detector coordinates in the background 
field as background and the residual background as a second 
model component with model parameters fixed to the values 
found in step (i). The overall spectra were fitted by a 
``mekal'' model (Mewe et al. 1985; Mewe et al. 1986; Kaastra 1992;
Liedahl et al. 1995; Arnaud \& Rothenflug 1985;
Arnaud \& Raymond 1992) with Galactic absorption 
(Dickey \& Lockman 1990; $n_{\rm H}=4.23 \times
10^{20} \; {\rm cm}^{-2}$). The results with and without 
introducing a residual background are consistent within 1 $\sigma$, e.g. the 
temperature (metallicity) of the $r<3^{\prime}$ region 
are $3.52\pm 0.17$ ($0.22\pm0.07$) and $3.61\pm 0.28$ ($0.22\pm0.11$).
  \begin{table} { \begin{center} \footnotesize
  {\renewcommand{\arraystretch}{1.3} \caption[]{Parameters of the
  residual background model fitted in the 0.4--15~keV band.  
  Col. (1): Instrument. Cols. (2--3):
  Index and normalization of the power law residual
  background in which the normalization was 
  scaled to 1~arcmin$^2$ in units of $10 ^{-4}$
  {$\rm cts~s^{-1}~keV^{-1}~arcmin^{-2}$}.} 
  \label{t:powb}}
  \begin{tabular}{lcc}
\hline   
\hline     
Instrument       & Index & Normalization \\ 
\hline
MOS1 &   1.95  & 1.00   \\
MOS2 &   2.21  & 0.64   \\
pn   &   1.95  & 2.25  \\
\hline  
\hline  
  \end{tabular}
  \end{center}
\hspace*{0.3cm}{\footnotesize 
}
  }
  \end{table}

In the image analysis for the surface brightness we carry out 
a vignetting correction for CDFS and CL0024$+$17, respectively. 
We derive the CDFS surface brightness in the same detector coordinates
as CL0024$+$17 and scale it using the ctr ratio of CL0024$+$17
and CDFS in the 10--12~keV (12--14~keV) band for MOS (pn). 
After subtracting the scaled CDFS as a background from CL0024$+$17,
we use a constant model to fit the residual soft X-ray background
which is quite flat in the outer region ($5<r<10^{\prime}$).  

For comparison, we also used the blank sky background from 
Lumb et al. (2002) and found that the parameters 
change less than 1 $\sigma$ for both spectral and surface brightness analyses.

\subsection{Point source subtraction}

In a preliminary spectral analysis without subtracting the point sources 
we found contamination in both the soft and 
hard bands, especially for pn. 

Therefore we generated a list of the bright point sources 
using SAS task
``edetect\_chain'' applied to five energy bands:
0.3--0.5~keV, 0.5--2~keV, 2--4.5~keV, 4.5--7.5~keV, 7.5--12~keV.
The sources detected include fourteen of sixteen 
sources which were found in ROSAT HRI observations 
(Soucail et al. 2000) except that S14 and S15 are out of the FOV of XMM-Newton.
Since we carried out analyses only in the $r<8^{\prime}$ region, 
only the brightest fourteen 
point sources within this region (including nine ROSAT HRI sources,
S1, S2, S3, S4, S5, S6, S8, S9 and S10 in Soucail et al. 2000)  
were subtracted from the source events using a 
radius of $40^{\prime\prime}$, which is comparable to the 
XMM-Newton Point Spread Function (PSF) cutoff radius of 
$\sim 45^{\prime\prime}$, for each source
that encloses nearly 90\% of the flux of the point sources in both the 
spectroscopic and image analyses (cf. Table~\ref{t:pointsrc}). 
\begin{table}
 { \begin{center} \footnotesize
  {\renewcommand{\arraystretch}{1.3} \caption[]{Parameters 
  for the brightest point sources detected in the $r<8^{\prime}$ region.
  Col. (1): No. of the point source. 
  Cols. (2--3): Central position.
  Col. (4): Flux determined out to a radius of $20^{\prime\prime}$
  in the 0.3--12~keV band. 
 } 
 \label{t:pointsrc}}
 \begin{tabular}{cccc}
 \hline
 \hline
No.   &  ${\rm R.A.}$ &  ${\rm decl.}$ & Flux   \\
      &           &           & ($10^{-13}~cgs$)\\
 \hline
1     &  00 26 12.8 &  17 03 46.8 &  1.317\\    
2     &  00 26 30.9 &  17 10 14.9 &  0.933\\
3     &  00 27 00.0 &  17 04 22.4 &  0.890\\
4     &  00 26 31.0 &  17 16 54.2 &  0.760\\
5     &  00 26 20.1 &  17 17 03.1 &  0.677\\
6     &  00 27 05.3 &  17 06 40.6 &  0.671\\
7     &  00 26 44.2 &  17 02 29.4 &  0.581\\
8     &  00 26 26.1 &  17 09 37.3 &  0.558\\
9     &  00 27 07.5 &  17 07 48.6 &  0.542\\
10    &  00 26 17.2 &  17 03 06.1 &  0.400\\
11    &  00 26 45.8 &  17 12 30.5 &  0.391\\
12    &  00 26 17.9 &  17 09 45.8 &  0.259\\
13    &  00 27 03.7 &  17 07 21.5 &  0.224\\
14    &  00 26 42.8 &  17 08 30.9 &  0.060\\
 \hline
 \hline
 \end{tabular}
 \end{center}
\hspace*{0.3cm}{\footnotesize 
}  } \end{table}

\section{Results}
\label{s:result}

\subsection{Image analysis}

The images in the 0.5--2~keV band for MOS and pn were  
corrected for vignetting using their weighted exposure maps
and were then combined.
We created an adaptively-smoothed image from the combined image
with a maximum sigma (the width of the smoothing Gaussian) 
of $12^{\prime \prime} \times 12^{\prime \prime} $  
and a signal-to-noise of 5.0
(see Fig.~\ref{f:image}).
Superposed XMM-Newton X-ray contours indicate an elongation in 
the northwest-southeast (NW-SE) direction and substructure in the NW 
(${\rm R.A.}=00^h26^m25^s.8$, 
${\rm decl.}=17^{\circ}12^{\prime}03^{\prime\prime}.7$, 
$\sim 3.3^{\prime}$ from the center).

We found no sign of the substructure described 
in Bonnet et al. (1994) which is $\sim 6.9^{\prime}$
from the cluster center
(${\rm R.A.}=00^h26^m35^s.7$, ${\rm decl.}=17^{\circ}09^{\prime}35^{\prime\prime}.8$)
to the northeast (NE). However, we observed the substructure, 
described in Kneib et al. (2003) and Czoske et al. (2001; 2002), located  
$\sim 3.3^{\prime}$ NW from the cluster center
covering $1^{\prime}$ with a temperature of $0.87 \pm 0.13$~keV fixing 
the metallicity to $Z=0.3Z_\odot$ and a bolometric X-ray luminosity of
$L_{\rm X}^{\rm bol}=0.57 \pm 0.06\times 10^{43} h^{-2}_{70}$~erg~s$^{-1}$, 
2\% of the total emission of the cluster.
We also observed emission from a pair of background groups in the NE 
which were decribed in Czoske et al. (2002) at $z=0.495$ based on a 
wide-field spectroscopic survey.
The northern group centered at 
${\rm R.A.}=00^h26^m50^s.1$, ${\rm decl.}=17^{\circ}19^{\prime}37^{\prime\prime}.8$ 
is more compact and X-ray luminous 
($L_{\rm X}^{\rm bol}=0.24\times 10^{44} h^{-2}_{70}$~erg~s$^{-1}$, 
$r<100^{\prime \prime}$).
The southern group centered at 
${\rm R.A.}=00^h27^m00^s.9$, ${\rm decl.}=17^{\circ}14^{\prime}51^{\prime\prime}.7$
is too faint for spectral analysis. If
we assume a gas temperature in the range 1--4~keV, then the bolometric 
luminosity of this group lies in the range 
0.076--0.079~$ \times 10^{44} h^{-2}_{70}$~erg~s$^{-1}$ 
($r<100^{\prime \prime}$).

\subsection{Chandra observation of the center}

CL0024$+$17 was also observed by Chandra (ID: 929).
Ota et al. (2004) showed an adaptively smoothed image of the cluster
with a rather symmetric appearance. 

A closer inspection reveals asymmetries, however.
Fig.~\ref{f:cdf} shows an unsmoothed image
(0.5--2.5~keV, $\sim$2$^{\prime \prime}$ pixels) and 
surface brightness profiles in annuli using the 0.5--2.5~keV image 
with azimuths 30$^{\circ}$--120$^{\circ}$ (counter clockwise) 
from north to east and from south to west, respectively. 
There is a clear asymmetry between the northeast and southwest. 
To the northeast, we note a sharp decrease at a radius of about 
$34^{\prime\prime}$. The profile to the southwest displaying a 
more gradual decline. In Fig.~\ref{f:ktZ}, we show the azimuthally averaged 
Chandra temperature profile from a more detailed spectral study (kindly 
provided by Alexey Vikhlinin) fixing the metallicity to $Z=0.3Z_\odot$.
We note a corresponding temperature decrease approximately at the 
same radius as the surface brightness decline. 
This complex structure including a possible indication of a ``shock front''
might provide further support to the cluster merger scenario.

\subsection{Optical observations}

To compare the X-ray morphology with the optical image in 
the central region of CL0024$+$17,
we retrieved 15 HST WFPC2 observations including  
11 from PI: Ellis (ID: 8559) and
4 from PI: Turner (ID: 5453).
They are block 0, 11, 12, 13, 18, 19, 23, 24, 25, 36, 
37 and 38 in Treu et al. (2003).
All images were taken in the F814W filter (I band).
We use the IRAF STSDAS and IMAGES packages to obtain the mosaic image 
from the calibrated science images.

The X-ray contours of the XMM-Newton observations were plotted  
on the merged HST image (shown in Fig.~\ref{f:hstimage}).
We note the overlapping cluster
centers and galaxy concentrations in the NW ($\sim 3.3^{\prime}$)
of the X-ray and optical images. 

\subsection{Hardness ratio maps}

To further study the substructure in CL0024$+$17
we produced hardness ratio map (HRM) of the cluster, 
which primarily reflects the temperature distribution. 
The hardness ratio is the photon flux ratio of the hard band of 2--7.5~keV 
to the soft band of 0.5--2~keV, $HR=I$(2--7.5~keV)$/I$(0.5--2~keV). 
An indication of an elongation in the NW-SE direction is visible
in the HRM (see Fig.~\ref{f:hnr}) on a scale of $r \sim 3.3^{\prime}$.
The HRM shows a complicated structure with an average $HR$ of $0.55\pm0.39$  
in the central $r < 3^{\prime}$ region. The substructure
in the NW ($\sim 3.3^{\prime}$) has an average of $HR=0.48\pm 0.34$ 
within $1^{\prime}$. We note two peaks: one is at 
(${\rm R.A.}=00^h26^m24^s.8$, ${\rm decl.}=17^{\circ}11^{\prime}13^{\prime\prime}.2$) 
that is marginally coincident with the NW substructure; the other is close 
to the center at 
${\rm R.A.}=00^h26^m30^s.9$, ${\rm decl.}=17^{\circ}09^{\prime}50^{\prime\prime}.3$.

\subsection{Temperature and metallicity profiles}
\label{s:tem}

We first derive global X-ray properties of
CL0024$+$17 from the spectra
extracted from the 
$r<3^{\prime}$ region, covering radii up to a spherical 
overdensity of $\sim 250$, i.e. the ratio of
the mean density of the dark halo with 
respect to the redshift-dependent critical density
$\rho_{\rm crit}(z)$.
We measure an emission weighted temperature of 
$3.52\pm 0.17$~keV, metallicity of $0.22\pm0.07$ and 
bolometric luminosity of 
$2.9 \pm 0.1 \times 10^{44} h^{-2}_{70}$~erg~s$^{-1}$ 
($\chi^2/dof=248.5/228$).
We confirm the 
estimates of 3--4.5~keV and 
$L_{X}^{bol}=2.40\times 10^{44} h^{-2}_{70}$~erg~s$^{-1}$ 
in B\"ohringer et al. (2000) from the ROSAT observation 
and a estimate of $5.7^{+4.2}_{-2.1}$~keV 
in Soucail et al. (2000) from the ASCA observations.
The Chandra observations yield a comparable temperature 
of $4.47^{+0.42}_{-0.27}$~keV and a bolometric luminosity 
of $2.60\times 10^{44} h^{-2}_{70}$~erg~s$^{-1}$ (Ota et al. 2004). 
The source spectra for $r<3^{\prime}$       
and the fitted model are shown in Fig.~\ref{f:tempha}.
A lower luminosity of $1.26\times 10^{44} h^{-2}_{70}$~erg~s$^{-1}$ 
is predicted from the $L$--$T$ relation in Arnaud \& Evrard (1999) and
the XMM-Newton temperature measurement. The high luminosity 
might indicate the complex cluster center which will be discussed in
Sect.~\ref{s:discussion}.

For a more detailed spectral study we
divide the cluster region into five annuli: 0--0.4$^{\prime}$,
0.4--0.8$^{\prime}$, 0.8--1.3$^{\prime}$, 1.3--2$^{\prime}$, 
and 2--3$^{\prime}$.
We investigate the dependence of the temperature and 
metallicity measurements on the low energy (low-E) cut-off
(0.4~keV or 1~keV) as we did in our earlier study (Zhang et al. 2004a)
to test the robustness of the spectral data fitting.
We find that 
in the $r<1^{\prime}$ region the derived temperatures increase
by 15\% if the low-E cut-off is increased from 0.4~keV to 1~keV.
With the decreasing signal-noise-ratio (S/N) in the $r>1^{\prime}$ region
the derived temperatures increase by up to 40\% 
if the low-E cut-off is increased from 0.4~keV to 1~keV. As in our 
earlier work (see also Pratt \& Arnaud 2002) we attribute this change 
in the best fit temperature with energy range to the influence of a 
contaminating soft spectral component. To minimize its effect,
we use the higher energy cut-off at 1~keV above which the results are
stable.

The radial metallicity distribution is very flat which is expected
from simulations where merging efficiently flattens the metallicity profile 
(Kobayashi 2003). 
The deprojection correction does not provide a significant 
change in the results.
In the $r<1^{\prime}$ ($r>1^{\prime}$) region the temperature measurements 
vary by a factor of 5\% (50\%), which is approximately 
within the error bars. The temperature and metallicity 
profiles determined from the 
spectra in the 5 annuli 
($\chi^2/dof=68.2/72,77.0/78,64.0/60,50.2/60,110.1/96$ 
from the inner to the outer annulus) 
are shown in Fig.~\ref{f:ktZ}.
A variation of the temperature with radius is found 
here for the first time for CL0024$+$17.
The inner three bins covering the radial range to a radius of 
$1.3^{\prime}$ ($416 h^{-1}_{70}$~kpc) are consistent with
an isothermal profile. This agrees 
with the Chandra results by Ota et al. (2004) who found an isothermal 
temperature profile to an outer radius of $1.5^{\prime}$ 
($480 h^{-1}_{70}$~kpc). 
The form of the temperature profile is very similar to the profile 
we obtained for the sample of REFLEX-DXL 
(ROSAT-ESO Flux-Limited 
X-ray cluster survey, Distant X-ray Luminous)
clusters (Zhang et al. 2004b).
We scaled the temperature profile in Markevitch et al. (1998) 
to CL0024$+$17 using a radius of $r_{180}=1.10 h^{-1}_{70}$~Mpc 
(obtained in Sect.~\ref{s:mass}) and  
an emission weighted global temperature of 3.52~keV. 
We find good overall agreement as found for the REFLEX-DXL 
clusters except that the flat part of the temperature profile extends
to a larger radius than the average profile of Markevitch et al. (1998).
We also scaled the temperature profile for Sersic~159$-$03 in Kaastra et al. 
(2001) to CL0024$+$17, and found a similarly rapid temperature drop
as a function of radius as CL0024$+$17.

We found an isothermal temperature of $\sim 3.9$~keV to a radius of 
$1.5^{\prime}$ ($480 h^{-1}_{70}$~kpc) and a power law model with an index 
$\gamma=0.98$ outside this radius.
We approximate the temperature 
profile with two analytical models:
(a) an isothermal model using the global temperature of $3.52\pm0.17$~keV; 
(b) a function with the shape of a Gaussian of 
$k_{\rm B}T(r)=k_{\rm B}T_{\rm g} \cdot e^{ -\frac{1}{2} 
(\frac{r-r_{\rm g}}{w})^2}$ 
(see Table~\ref{t:fit}) which approximately fits the 
observed temperature profile. 

\subsection{Surface brightness}
\label{s:sx}

We obtained an azimuthally averaged surface brightness profile
for CL0024$+$17 in the 0.5--2~keV band for each instrument, 
in which the central position is 
${\rm R.A.}=00^h26^m35^s.7$, ${\rm decl.}=17^{\circ}09^{\prime}35^{\prime\prime}.8$.
This energy band is selected because it provides an almost 
temperature-independent X-ray emission coefficient over the 
expected temperature range. 
After the vignetting correction and double-step 
background subtraction described above, 
we find that a $\beta$-model (e.g. Cavaliere \& Fusco-Femiano 1976; 
Jones \& Forman 1984) 
\begin{equation}  
S_{x}(r)=S_0(1+\frac{r^2}{r_{\rm c}^2})^{-3\beta+1/2}
\end{equation}  
convolved with the XMM-Newton PSF
provides an adequate $\chi^2$ fit to the surface brightness profile 
(Fig.~\ref{f:sx}). A PSF at each position for each instrument is determined 
from an empirical calibration (Ghizzardi 2001) according to 
a mean energy $\sim 1.25$~keV of the 0.5--2~keV band 
and its off-axis radius.

We compare the PSF convolved and unconvolved $\beta$-model parameters  
in Table~\ref{t:beta}, and find that the core radii are overestimated 
and the slope parameters are underestimated in the 
latter case. The combined 90\% confidence contours 
(see Fig.~\ref{f:contour}) of the three instruments provide
a narrow range for $r_{\rm c}$ and $\beta$ 
($0.266<r_{\rm c}<0.295^{\prime}$; $0.551<\beta<0.593$)
at the 90\% confidence level.
B\"ohringer et al. (2000) obtained 
$r_c=0.11$--$ 0.28^{\prime}$ and $\beta=0.425$--$0.550$
from the ROSAT HRI data corrected for PSF. 
Ota et al. (2004) obtained 
$r_c=0.287\pm0.014^{\prime}$and $\beta=0.55\pm0.02$
from the Chandra observations.
Our result confirms the results from previous X-ray observations 
with a higher accuracy 
because of the improvement of the image quality of XMM-Newton 
compared to previous observations.

The surface brightness profile approximated by a $\beta$-model
can be analytically deprojected to yield the emission per volume
element, $\xi (r) = \widetilde{\Lambda}(r) n^2_{\rm e}(r)$. 
With the given emissivity, $\widetilde{\Lambda}(r)$, 
from the applied plasma model, one can derive the electron 
density profile 
$n_{\rm e}(r)=n_{\rm e0}(1+\frac{r^2}{r_{\rm c}^2})^{-3\beta/2}$ 
with the parameters given 
in Table~\ref{t:beta} combining the three detectors.
We studied the influence of the central emission in the
beta model analysis and found that the parameters like 
$\beta$ and $r_{\rm c}$ change within 2\%  
if the central bins are masked.
  \begin{table*} 
{ 
 \begin{center} \footnotesize
{
\renewcommand{\arraystretch}{1.3} \caption[]
{
  Parameters of the
  $\beta$-model.  
  Col. (1): Instrument. Col. (2): ``Y'' or ``N'' means 
  the PSF convolved or unconvolved $\beta$-model.
  Col. (3): Central surface brightness 
  (0.5--2~keV) in cts~s$^{-1}$~arcmin$^{-2}$. 
  Col. (4): Core radius in arcmin. Col. (5): Slope parameter. 
  Col. (6): $\chi^2/dof$.}
  \label{t:beta}
}
  \begin{tabular}{lccccl}
\hline   
\hline     
 & PSF & $S_0$ & $r_c$ & $\beta$ & $\chi^2/dof$\\ 
\hline
MOS1 &   Y   &  $0.106 \pm 0.015$ & $0.27 \pm 0.03$  & $0.57 \pm 0.02$ & $99.7/63$\\
     &   N   &  $0.047 \pm 0.004$ & $0.34 \pm 0.03$  & $0.56 \pm 0.01$ & $204.3/150$\\ 
MOS2 &   Y   &  $0.094 \pm 0.010$ & $0.32 \pm 0.03$  & $0.58 \pm 0.02$ & $79.2/65$\\
     &   N   &  $0.045 \pm 0.003$ & $0.36 \pm 0.03$  & $0.54 \pm 0.01$ & $159.7/150$\\ 
pn   &   Y   &  $0.375 \pm 0.033$ & $0.25 \pm 0.02$  & $0.56 \pm 0.01$ & $130.5/88$\\
     &   N   &  $0.158 \pm 0.007$ & $0.33 \pm 0.02$  & $0.54 \pm 0.01$ & $206.3/150$\\ 
\hline  
\hline  
  \end{tabular}
  \end{center}
\hspace*{0.3cm}
{
\footnotesize 
}
}
  \end{table*}

We obtain the pressure distribution $P(r)$ as a function of radius
directly from the temperature and electron density measurements.
Thus a $\beta$-model is applied to fit the pressure with results shown in 
Table~\ref{t:fit}.

Furthermore, we simulated a symmetric, background subtracted 
and flat fielded image using 
the parameters of the $\beta$-model in Table~\ref{t:beta}.
The residual map, which contains the background, 
is obtained by extracting 
the simulated image from the adaptively-smoothed image.
We confirm the existence of excess emission in the substructure in the NW, 
while the negative residual surface brightness in the cluster center
shows that the elongation flattens
the surface brightness compared to a symmetric structure.

To further test if the cluster mass profile can be described 
by a NFW profile (Navarro et al. 1997; NFW), we fit the observed 
surface brightness profile by the model computed from 
the NFW model described dark matter halo (e.g. Makino et al. 1998).
A model that fits the data well over a large range of radii shows
an inner cusp in the surface brightness. The central surface brightness 
model at the given resolution is about a factor of two to three higher 
and thus inconsistent with our observations. Since we detect signatures 
of a cluster merger in the central region. We are not surprised to find 
deviations from the NFW model. Also Tyson et al. (1998) found a core in the 
mass profile in their lensing analysis.  

\subsection{Cooling time of the gas}

The cooling time is the total energy of the gas divided 
by the energy loss rate (e.g. Zhang \& Wu 2003)
\begin{equation}
t_{\rm c}=2.869 \times 10^{10}  {\rm yr}\;\left
( \frac{1.2}{g} \right )
\left
(\frac{k_{\rm B} T}{\rm keV}
\right )^{\frac{1}{2}}\left (\frac{n_{\rm e}}{10^{-3} {\rm cm^{-3}}}
 \right )^{-1},
\label{eq:tcool}
\end{equation}
where $n_{\rm e}$ and $k_{\rm B}T$ are the electron number density and
temperature,
respectively, and $g$ is the Gaunt factor (a function of $k_{\rm B}T$).
The resulting cooling time as a function of radius is shown in
Fig.~\ref{f:tc}. The central cooling time of 4.5~Gyr is
smaller than the age of the Universe and probably smaller than the 
age of the cluster if we assume that the 
age of the cluster mass concentraion is longer than two crossing time 
because the merger frequency is expected to be 
a few Gyr (e.g. Schuecker et al. 2001). 
An even smaller central cooling time is implied by the Chandra data 
(Ota et al. 2004) due to the higher angular resolution of 
the central region. The data thus correspond, in the classical 
interpretation, to a small or moderate cooling flow (Fabian \& Nulsen 1977).

\subsection{Gas entropy}

The main part of the observed entropy, defined as 
$S=k_{\rm B}T n_{\rm e}^{-2/3}$ (e.g. Ponman et al. 1999), 
results from shock heating of 
the gas during cluster formation and it scales with the cluster
critical temperature. An excess above this scaling law indicates the effect 
of an additional, non-gravitational heating source 
(e.g. Lloyd-Davies et al. 2000).
While a low central entropy indicates significant 
radiative cooling. We show the entropy profile derived for 
CL0024$+$17 from the XMM-Newton observations in Fig.~\ref{f:scaleds}.
 
The entropy of the $r>0.1 h^{-1}_{70}$~Mpc region 
derived from temperature measurements 
of CL0024$+$17 has a similar slope as
that predicted from a spherical accretion shock model, 
$S\propto r^{1.1}$ (Kay 2004; thick grey line in Fig.~\ref{f:scaleds}). 
The entropy under the assumption of isothermality becomes steeper
than the entropy derived from temperature measurements, 
especially in the outer region.

The entropy of the gas in the center ($r<0.1h^{-1}_{70}$~Mpc) lies below the 
entropy floor, $S\sim 124 h^{-1/3}_{70}$~${\rm keV~cm^2}$, 
derived for clusters by Lloyd-Davies et al. (2000).
This indicates some effect from radiative cooling. 

We scale the entropy of CL0024$+$17 by 
$1+\frac{k_{\rm B}T(r)}{k_{\rm B}T_0}$, where
$k_{\rm B}T_0=2$~keV is a constant related to the degree of preheating.
In Fig.~\ref{f:scaleds}, we compare it to 
the entropy of the Birmingham-CfA clusters in the temperature
range of 2.9--4.6~keV (Ponman et al. 2003) scaled also by 
$1+\frac{k_{\rm B}T(r)}{k_{\rm B}T_0}$. 
The entropy of CL0024$+$17 agrees 
with the self-similarly scaled entropy 
described by Ponman et al. (2003) within the observed dispersion of
their cluster sample. 

Eq.(\ref{eq:tcool}) can be rewritten as 
\begin{equation}
\begin{small}
S_c = 100{\rm keV  cm^2} 
 	\left(\frac{t_{\rm c}}{2.869\times 10^{10}{\rm yr}}
	\right)^{\frac{2}{3}}
	\left(\frac{g}{1.2}\right)^{\frac{2}{3}} 
	\left (\frac{k_{\rm B}T}{\rm keV}\right)^{\frac{2}{3}}.
\end{small}
\end{equation}
If the cooling time $t_{\rm c}$ is chosen approximately to be 
the age of the Universe,
the above equation would allow us to 
determine the critical entropy floor $S_{\rm c}$ of the gas 
in the cluster before the onset of significant cooling and possible 
mass deposition. In the center of CL0024$+$17 the observed entropy 
is slightly below this value ($\sim$~154~keV~cm$^2$)
which implies a insignificant radiative cooling.

\subsection{Mass modeling} 
\label{s:mass} 

We assume the intracluster gas to be in hydrostatic equilibrium within
the gravitational potential dominated by dark matter. 
Neglecting the cosmological constant, $\Lambda$,
at these high overdensities 
(a less than 1\% effect, see Zhang et al. 2004a),
we have
\begin{equation}
\frac{1}{\mu m_{\rm p} n_{\rm e}}\frac{d(n_{\rm e} k_{\rm B} T)}{dr}=
  -\frac{GM(r)}{r^2}~, 
\label{e:hyd}
\end{equation}
where $\mu=0.62$ is the mean
molecular weight per hydrogen atom.

Analytic models of the gas density and pressure profiles derived above can be
easily combined with Eq.(\ref{e:hyd}) to obtain the mass profile
(Fig.~\ref{f:mass}). 
The virial radius (not $r_{200}$) is defined to be 
a radius with overdensity (the average density with respect to 
$\rho_{\rm crit}$ at the cosmic epoch $z$) of
$\Delta_{\rm c}=18\pi^2+82[\Omega_{\rm M}(z)-1]-39[\Omega_{\rm M}(z)-1]^2$ 
for a flat universe. $\Omega_{\rm M}(z)$ is the cosmic density 
parameter (e.g. Zhang \& Wu 2003).
Therefore the mass profiles in Fig.~\ref{f:mass} 
combined with the expression for the overdensity yield 
$r_{\rm vir}=1.37 h^{-1}_{70}$~Mpc using the global temperature
and $r_{\rm vir}=1.26 h^{-1}_{70}$~Mpc using the observed temperature 
distribution.

We compare the XMM-Newton results with the recent measurements
from Chandra and HST in Table~\ref{t:masscom} and Fig.~\ref{f:mass}.
The XMM-Newton measurements using the measured temperature profile
are slightly lower than Chandra values derived under  
the assumption of isothermality.
As shown in Table~\ref{t:masscom} and Fig.~\ref{f:mass}, 
the discrepancy between the X-ray measured total mass 
and strong lensing mass at the radii of $0.143h^{-1}_{70}$~Mpc 
(Broadhurst et al. 2000) and $0.153h^{-1}_{70}$~Mpc (Tyson et al. 1998)
remains large, about a factor of 4. 

  \begin{table*} { \begin{center} \footnotesize
  {\renewcommand{\arraystretch}{1.3} \caption[]{Comparison of the
	gravitational mass from the 
	X-ray and optical lensing measurements.
}  \label{t:masscom}}
  \begin{tabular}{lllllll}
\hline   
\hline     
\multicolumn{2}{c}{ Observation} & \multicolumn{2}{c}{XMM-Newton $^a$} & Chandra $^b$ & \multicolumn{2}{c}{HST}    \\
\hline  
$r_{200}$ & ($h^{-1}_{70}$~Mpc)         &   1.11 &  1.05 & 1.4 & 1.7 & $^c$\\
$M_{200}$ & ($10^{14} h^{-1}_{70}{\rm M_{\odot}}$)	   & $2.3 \pm 0.1$ & $2.0 \pm 0.3$ & $4.6^{+0.7}_{-0.5}$
		& $5.7^{+1.1}_{-1.0}$ & $^c$\\
$M_{\rm proj}(r=0.143)$ & ($10^{14} h^{-1}_{70}{\rm M_{\odot}}$)  & $0.61 \pm0.01$  & $0.48 \pm0.03$ & --- & $1.59 \pm 0.04$
		& $^d$\\
$M_{\rm proj}(r=0.153)$ & ($10^{14} h^{-1}_{70}{\rm M_{\odot}}$)  &  $0.64 \pm 0.01$& $0.51 \pm 0.03$& $^{0.60^{+0.14}_{-0.09}}_{0.53^{+0.13}_{-0.08}}$ $^{(\beta)}_{\rm (NFW)} $ & $ 2.37 \pm 0.36$
		& $^e$\\
$M(r=1.4)$ & ($10^{14} h^{-1}_{70}{\rm M_{\odot}}$)         & $3.0 \pm 0.2$ & $2.4 \pm 0.3$   & $4.6^{+0.7}_{-0.5}$ & ---
		& \\
$M(r=1.7)$ & ($10^{14} h^{-1}_{70}{\rm M_{\odot}}$) 	     &  $3.6 \pm 0.2$& $2.8 \pm 0.4$& --- & $5.7^{+1.1}_{-1.0}$
		& $^c$\\
\hline   	
\hline  
  \end{tabular}
  \end{center}
\hspace*{0.3cm}{\footnotesize}
$^a$ XMM-Newton results using
the global temperature and using the observed temperature profile in which 
projected masses use the truncation radius of $r_{200}=2.5 h^{-1}_{70}$~Mpc;
$^b$ Ota et al. (2004); $^c$ Kneib et al. (2003); 
$^d$ Broadhurst et al. (2000); $^e$ Tyson et al. (1998).\\
}
  \end{table*}

Under the assumption of isothermality, the XMM-Newton measurements are more 
consistent with the observationally determined $M_{200}$--$T$ relation,
\begin{equation}
M_{200}= 4.11 \pm 0.02 \times 10^{13} h_{70}^{-1} \left (
\frac{kT_{\rm B}}{\rm keV} \right )^{1.60\pm 0.04}{\rm M_{\odot}}\; ,
\label{e:mt}
\end{equation}
which is based on the conventional $\beta$ model for the X-ray surface 
brightness profile and hydrostatic equilibrium for 22 nearby 
clusters from Xu et al. (2001). For CL0024$+$17, Eq.(\ref{e:mt}) provides 
$M_{200} = 3.08 \times 10^{14}$~$h_{70}^{-1}{\rm M_{\odot}}$ using the 
X-ray temperature. 
The masses obtained from the X-ray temperature measurement 
via the $M$--$T$ relations in Finoguenov et al. (2001)
and in Bryan \& Norman (1998) are
$M_{200} = 5.96 \times 10^{14}$~$h_{70}^{-1}{\rm M_{\odot}}$
and $M_{200} = 3.76 \times 10^{14}$~$h_{70}^{-1}{\rm M_{\odot}}$
in which the former is more consistent with 
the HST optical lensing estimate.

Navarro et al. (1997; NFW) describe a universal density profile from
numerical simulations in hierarchical clustering scenarios
\begin{equation}
\rho_{\rm DM}(r)=\delta_{\rm crit}\rho_{\rm crit}\left(\frac{r}{r_s} \right)^{-1}\left(1+ \frac{r}{r_s}\right)^{-2}~, 
\label{e:nfw}
\end{equation}
where $\delta_{\rm crit}$ and $r_s$ are the characteristic density and
scale of the halo, respectively, $\rho_{\rm crit}$ is the critical
density of the Universe at the cosmic epoch $z$ and 
$\rho_{\rm s}=\delta_{\rm crit}\rho_{\rm crit}$.  $\delta_{\rm crit}$
is related to the concentration parameter of a dark halo $c=r_{\rm
vir}/r_s$ by
\begin{equation}
\delta_{\rm crit}=\frac{200}{3}c^3 \left [\ln(1+c)-\frac{c}{1+c} \right ]^{-1}~.
\label{e:nfwc}
\end{equation}

We fit the X-ray determined mass profile with a NFW model,
and found that the NFW model does not provide a good fit to our data 
in the $r<0.1h^{-1}_{70}$~Mpc region (Fig.~\ref{f:mass}). 
However the NFW model fit provides a virial radius of 
$c r_s=1.28 h^{-1}_{70}$~Mpc which is consistent with the virial 
radii of $r_{\rm vir}=1.37 h^{-1}_{70}$~Mpc and 
$r_{\rm vir}=1.26 h^{-1}_{70}$~Mpc 
derived from the mass profiles using the global temperature and 
using the observed temperature distribution, respectively, 
in Fig.~\ref{f:mass} combined with the 
expression for the overdensity. We derive a radius of 
$r_{200}=0.93 h^{-1}_{70}$~Mpc from the empirical relation $r_{200}=0.813 \; 
(kT_{\rm B}/{\rm keV})^{0.5}(1+z)^{-3/2}h^{-1}_{70}$~Mpc
(Navarro et al. 1995), which is slightly lower than 
$r_{200}= 1.11 h^{-1}_{70}$~Mpc using the global temperature and 
$r_{200}=1.05 h^{-1}_{70}$~Mpc using the observed temperature distribution. 

Furthermore, we apply an extended NFW model 
(e.g. Hernquist 1990; Zhao 1996; Moore et al. 1999), 
\begin{equation}
\rho_{\rm DM}(r)=\delta_{\rm crit}\rho_{\rm crit}
\left (\frac{r}{r_s} \right )^{-\alpha} \left (1+\frac{r}{r_s} \right )^{\alpha-3} 
\label{e:exnfw}
\end{equation}
to fit the data (Table~\ref{t:fit}) which provides a negative index. 
This indicates that the dark 
matter density has a flat core for CL0024$+$17. 

  \begin{table} { \begin{center} \footnotesize
  {\renewcommand{\arraystretch}{1.3} \caption[]{Parameters of each profile
  model for the best $\chi^2$ fit.}  \label{t:fit}}
  \begin{tabular}{lll}
\hline   
\hline     
 Model        &  \multicolumn{2}{c}{parameter}    \\
\hline  
 Gaussian      & $k_{\rm B}T_{g}$ (keV)   & $4.2\pm 0.4$\\
	      & $r_{g}$ ($h_{70}^{-1}$Mpc)            & $0.08\pm 0.04$\\
	      & $w$     & $0.69 \pm 0.04$\\
 $\beta$      & $r^{n_{\rm e}}_{\rm c}$ ($h_{70}^{-1}$Mpc) 
		& $0.093\pm 0.001$ \\
              & $n_{\rm e0}$ ($10^{-3}{\rm cm}^{-3}$) 
		& $8.8 \pm 0.1$\\
              & ${\beta}^{n_{\rm e}}$ & $0.57 \pm 0.01$ \\
              & $r^{\rm P}_{\rm c}$ ($h_{70}^{-1}$Mpc) & $0.11\pm 0.01$ \\
              & $P_{\rm 0} (10^{-2}{\rm keV} {\rm cm}^{-3})$     
		& $3.59 \pm 0.04$\\
              & ${\beta}^{\rm P}$        & $0.68 \pm 0.01$ \\	 
 NFW          & $r_{s}$ ($h_{70}^{-1}$Mpc) & $0.37\pm 0.03$ \\
	      & $\rho_{s}$ ($10^{14}{\rm M_{\odot} Mpc}^{-3}$) 
		& $5.3 \pm 0.7$\\	
	      & c                              & $3.5$ \\
 extended NFW & $r_{s}$ ($h_{70}^{-1}$Mpc) & $0.072\pm 0.004$ \\
              & $\rho_{s}$ ($10^{16}{\rm M_{\odot} Mpc}^{-3}$) 
		& $4.3 \pm 1.7$\\
	      & $\alpha$ & $-0.8 \pm 0.1$ \\
\hline   	
\hline  
  \end{tabular}
  \end{center}
\hspace*{0.3cm}{\footnotesize}

}
  \end{table}

We also present a comparison of the projected mass density and 
projected mass using the truncation radius of 
either $r_{200}=1.7 h^{-1}_{70}$~Mpc,
predicted from the lensing data in Kneib et al.(2003)
or a larger radius of $r=2.5 h^{-1}_{70}$~Mpc (Fig.~\ref{f:mass}).
The discrepancy is larger between the X-ray and strong lensing 
measurements than between the X-ray and weak lensing 
measurements. 

\subsection{Gas mass fraction}

The gas mass fraction distribution according to the
definition $f_{\rm gas}(r)=M_{\rm gas}(r)/M(r)$ 
is shown in Fig.~\ref{f:fg}. We found 
$f_{\rm gas}=0.19 \pm 0.01 h^{-3/2}_{70}$ at $r_{200}=1.11h^{-1}_{70}$~Mpc 
using the global temperature
and $0.20\pm0.03 h^{-3/2}_{70}$ at $r_{200}=1.05 h^{-1}_{70}$~Mpc 
using the observed temperature profile, respectively.
This result is comparable with the Chandra results 
$f_{\rm gas}=
0.14^{+0.03}_{-0.02}h^{-3/2}_{70}$ at 
$r_{200}=1.40 h^{-1}_{70}$~Mpc (Ota et al. 2004),
and agrees with the WMAP measured baryon 
fraction of the Universe 
$f_{\rm b}=\Omega_{\rm b}/\Omega_{\rm
m}=0.166$, where $\Omega_{\rm b}~ h^2=0.0224$ and $\Omega_{\rm m}~
h^2=0.135$ (Spergel et al. 2003). 
We obtain a gas mass fraction of $0.11 \pm 0.01$ using the 
global temperature at $r_{2500}=0.30$~Mpc and 
$0.09 \pm 0.01$ using the observed temperature profile at $r_{2500}=0.33$~Mpc,
which agrees with the measurements of Allen et
al. (2002) based on Chandra observations of seven clusters yielding
$f_{\rm gas}\sim 0.105$--$0.138h^{-3/2}_{70}$.
However, this gas mass fraction at $r_{200}$ is 
slightly higher than the measurements of Sanderson
et al. (2003) based on ASCA/GIS, ASCA/SIS and ROSAT/PSPC observations of 66
clusters yielding $f_{\rm gas}=0.13\pm 0.01~h_{70}^{-3/2}$,
the measurements of Ettori et al. (2002) based on BeppoSAX 
observations of 22 nearby clusters, and the gas mass fraction 
for A1413 (Pratt \& Arnaud 2002) at 
$z=0.143$ based on XMM-Newton observations yielding
$f_{\rm gas}\sim 0.12h^{-3/2}_{70}$.

The global parameters based on the 
XMM-Newton observations for CL0024$+$17 are given in Table~\ref{t:global}.
  \begin{table} { \begin{center} \footnotesize
  {\renewcommand{\arraystretch}{1.3} \caption[]{
	Measured parameters for CL0024$+$17. 
  Col. (1): Parameter. Cols. (2-3): Values obtained: 
	(i) in the $r<3^{\prime}$ region for line 2--4;
	(ii) using the global temperature and observed 
	temperature distribution for line 5--7.
} 
  \label{t:global}}
  \begin{tabular}{lll}
\hline   
\hline     
 Parameter     & \multicolumn{2}{c}{Value}\\ 
\hline
$kT_{\rm B}$ (keV)      & \multicolumn{2}{c}{$3.52\pm0.17$}\\
$Z$ ({\rm $Z_{\odot}$}) & \multicolumn{2}{c}{$0.22\pm0.07$}\\
$L^{bol}_{X}$ ($10^{44} h^{-2}_{70}$~erg~s$^{-1}$) & \multicolumn{2}{c}{$2.9 \pm 0.1$} \\
$r_{200}$ ($h^{-1}_{70}$~Mpc) &   1.11 &  1.05 \\
$M_{200}$ ($10^{14} h^{-1}_{70}{\rm M_{\odot}}$)
	& $2.3 \pm 0.1$ & $2.0 \pm 0.3$ \\
$f_{\rm gas}$ ($h^{-3/2}_{70}$) & $0.19 \pm 0.01$ & $0.20\pm0.03$ \\
\hline  
\hline  
  \end{tabular}
  \end{center}
\hspace*{0.3cm}{\footnotesize 
}
  }
  \end{table}

\section{Discussion}
\label{s:discussion}

\subsection{Metallicity}

We measure a global metallicity of 
$0.22\pm0.07$~Z$_{\odot}$ which is typical 
for a cluster with an insignificant cooling flow (Fabian \& Nulsen 1977).
This metallicity value is in good agreement with the averaged 
metallicity $<Z>=0.21^{+0.10}_{-0.05}$~Z$_{\odot}$ for 18 distant 
clusters with redshift $0.3<z<1.3$ in Tozzi et al. (2003).
However, Ota et al. (2004) obtain a much higher metallicity 
of 0.61--0.95~Z$_{\odot}$.
We use the images in the 5--6~keV and 6--7~keV bands as indicators 
of the iron K line and continuum emission, respectively, 
and obtain an iron to continuum map. 
We extracted the spectra from a suggested 
high metallicity region in the iron-to-continuum map at 
${\rm R.A.}=00^h26^m32^s.7$, ${\rm decl.}=17^{\circ}09^{\prime}46^{\prime\prime}.3$
covering a radius of $0.32^{\prime}$.
No significant high metallicity 
was obtained by either setting a free temperature 
($5.4\pm 1.6$~keV) or setting 
the temperature to the emission weighted temperature (4.09~keV) 
of the zone covering radii of $0.4<r<0.8^{\prime}$.
For the same radius of $1.5^{\prime}$ as used by Ota et al. (2004),
we measure a metallicity of $0.23\pm 0.07$ and a temperature of 
$3.9\pm0.2$~keV (see Table~\ref{t:binkt}) using the XMM-Newton data.
Therefore the XMM-Newton data with higher significance provide no 
explanation for the high metallicity value obtained by Ota et al. (2004).
The difference between XMM-Newton and Chandra metallicity measurements 
is an about 2 sigma effect, and may be due to the limited photon statistics 
in the Chandra data.

\subsection{Temperature gradient}

We derive an isothermal temperature of $\sim 3.9$~keV to a radius of 
$1.5^{\prime}$ ($480 h^{-1}_{70}$~kpc) and a power law model with an index of 
$\gamma=0.98$ outside this radius.
The XMM-Newton temperature in the center is in 
good agreement with the Chandra measurement
which extends to 1.5$^{\prime}$ 
in the Chandra observation 
of CL0024$+$17, covering the almost isothermal center ($r<1.3^{\prime}$) 
in the XMM-Newton observations. 
We provide a more extended temperature profile
based on the XMM-Newton observations in which the  
temperature profile decreases in the
1.3--3$^{\prime}$ region.
The strong temperature gradient at large radii makes 
the gravitational mass lower by a factor of $\sim $~20--25\% compared to the 
isothermal results. 
We found that the temperature gradient is not a sudden drop at 
a radius of 1.3$^{\prime}$ by measuring the temperature in different 
bin sizes (see Table~\ref{t:binkt}).
  \begin{table} { \begin{center} \footnotesize
  {\renewcommand{\arraystretch}{1.3} \caption[]{Temperature and metallcity 
	measurements in difference bin size of the radius. 
}  \label{t:binkt}}
  \begin{tabular}{llll}
\hline   
\hline     
 Region & $k_{\rm B}T$ (keV) & $Z$ ({\rm $Z_{\odot}$}) & $\chi^2/dof$\\
\hline  
$r<0.40^{\prime}$  & $ 4.1\pm0.3 $ & $ 0.16 \pm 0.10$ & $68.2/72$\\
$r<0.50^{\prime}$  & $ 4.4\pm0.3 $ & $ 0.10 \pm 0.09$ & $69.1/69$ \\
$r<1.00^{\prime}$  & $ 4.1\pm0.2 $ & $ 0.25 \pm 0.08$ & $161.3/137$ \\
$r<1.25^{\prime}$  & $ 4.0\pm0.2 $ & $ 0.23 \pm 0.07$ & $159.6/152$ \\
$r<1.50^{\prime}$  & $ 3.9\pm0.2 $ & $ 0.23 \pm 0.07$ & $173.1/170$ \\
$r<1.67^{\prime}$  & $ 3.8\pm0.2 $ & $ 0.24 \pm 0.07$ & $170.9/163$\\
$r<2.00^{\prime}$  & $ 3.7\pm0.2 $ & $ 0.23 \pm 0.07$ & $185.3/184$ \\
$r<3.00^{\prime}$  & $ 3.5\pm0.2 $ & $ 0.22 \pm 0.07$ & $68.5/71$ \\
\hline   	
\hline  
  \end{tabular}
  \end{center}
\hspace*{0.3cm}{\footnotesize}
}
  \end{table}
Within a radius of $3^{\prime}$ for CL0024$+$17, 
we derived a polytropic index of 1.17 for the ICM in the outskirts
which is lower than the adiabatic index of 5/3.
This implies that the outskirts are 
still convectively stable.
We consider the uncertainty in the mass estimate caused by the temperature 
gradient by artificially increasing the slope parameter of 
the temperature distribution by a factor of $\sim$20\%, and 
find a decrease of the total mass by a factor of $\sim$10\%.  

Since the soft band (0.5--2~keV) is not so 
sensitive to the temperature map as the hard band (2--7.5~keV), 
the HRM can be used as an indicator of the temperature distribution. 
Therefore the temperature map can be obtained from the HRM
(e.g. Sanders \& Fabian 2001; Sanders et al. 2003; Akimoto et al. 2003). 
The $HR$ is also a reflection of absorption measure
which affects the temperature (e.g. Akimoto et al. 2003).
Thus we study the further properties of the features in the HRM
by performing a spectral analysis. For the statistic reason, 
we extracted the spectra from one of the peaks
suggesting high temperature in the HRM at
${\rm R.A.}=00^h26^m30^s.9$, ${\rm decl.}=17^{\circ}09^{\prime}50^{\prime\prime}.3$
covering a radius of $0.33^{\prime}$,
and obtained a temperature of 
$5.8\pm 1.0$~keV with metallcity fixed at 0.3~${\rm Z_{\odot}}$
and $5.7\pm 1.0$~keV with the metallicity free 
($Z=0.88^{+0.78}_{-0.58}\;{\rm Z_{\odot}}$, 
$L_{\rm X}^{\rm bol}=0.43\times 10^{44} h^{-2}_{70}$~erg~s$^{-1}$). 
This supports the indication in the HRM of a high 
temperature region and not 
an artifact due to metallicity variations. 
However, the spectra can also be 
fitted by a power law with an index of $1.74\pm 0.14$, which could then
be attributed to some contamination 
from an AGN. This hot region is located exactly at 
the position of the substructure described in Czoske et al. (2001; 2002).

\subsection{Complex structure}

The Chandra image of the cluster core displays
an indication of complex structure.
The projection of the counts as a function of radius using the high spatial 
resulotion image from Chandra shows an asymmetric structure 
and a surface brightness ``edge'' in the central region.
The galaxy alignment in the HST image is perpendicular to the ``edge''. 
Such an asymmetric central structure with an ``edge'' ($r<0.5^{\prime}$) 
in the XMM-Newton HRM support the idea of a cluster merger.

\subsection{Comparison to lensing measurements}

The strongest disagreement between the X-ray mass profiles 
and the strong lensing results is found in the cluster center. The difference 
between the X-ray determined mass (Table~\ref{t:masscom}) 
and the strong lensing mass of 
$1.59 \pm 0.04 \times 10^{14} h^{-1}_{70}{\rm M_{\odot}}$ 
at the arc radius of 
$0.143 h^{-1}_{70}$~Mpc (Broadhurst et al. 2000) and 
$ 2.37 \pm 0.36 \times 10^{14} h^{-1}_{70}{\rm M_{\odot}}$
at the arc radius of $0.153 h^{-1}_{70}$~Mpc (Tyson et al. 1998)
is up to a factor of $\sim 4$.

This discrepancy is most probably due to the disturbed structure 
of the cluster center, highlighted by the elongated X-ray 
iso-surface-brightness contours in the central region. 
The substructure found in the gravitational lensing mass 
distribution map by Kneib et al. (2003) is also supported by the X-ray 
image. The velocity distribution of the galaxies can approximately 
be explained by a line-of-sight merger of two systems with a mass ratio 
of the order of 1:2 (Czoske et al. 2001; 2002). Therefore, the application 
of hydrostatic equilibrium to determine the mass in the central region 
of the cluster from the gas properties may not be valid.

We observed an elongation in the NW-SE direction in the HRM.
The negative residual surface brightness in the cluster center
demonstrates that the elongation flattens 
the surface brightness compared to an azimuthally symmetric structure. 
The XMM-Newton image analysis shows substructure to the NW of the 
cluster center which is marked by a galaxy (see Fig.~\ref{f:hstimage}) 
in Kneib et al. (2003). 
This galaxy shows significantly extended X-ray emission  
compared to the XMM-Newton PSF and has a luminosity of 
$L_{\rm X}^{\rm bol}=0.57 \pm 0.06 \times 10^{43} h^{-2}_{70}$~erg~s$^{-1}$
and a temperature of $0.87 \pm 0.13$~keV within 1$^{\prime}$ at  
${\rm R.A.}=00^h26^m25^s.8$, ${\rm decl.}=17^{\circ}12^{\prime}03^{\prime\prime}.7$.
Since this high surface brightness region does not have a 
lower temperature but probably rather has a higher temperature
than its surroundings, it does not mark a ``cold front''
but a high pressure region which coincident with a mass concentration 
seen in the lensing map. This implies that the second mass 
peak is included within the gaseous halo of the cluster and not only
seen projected onto the cluster.

\subsection{Comparison to velocity dispersion measurement}

The $M_{200}$--$\sigma_{\rm DM}$ relation from simulations in 
Evrard \& Gioia (2002) gives 
a comparable velocity dispersion of $848^{+51}_{-53}$~km~s$^{-1}$ from 
the HST mass and a relatively low value of the velocity dispersion of
$598^{+29}_{-31}$~km~s$^{-1}$ using the XMM-Newton mass.
Girardi \& Mezzetti (2001) obtained a galaxy velocity dispersion  
of $911^{+81}_{-107}$~km~s$^{-1}$ for CL0024$+$17. 
The velocity dispersion derived from the XMM-Newton mass
is in good agreement with the velocity dispersion measure in 
Czoske et al. (2002) who found velocity dispersions of 
$561^{+95}_{-83}$ and $554^{+175}_{-304}$ for the central component 
and foreground component which suggests a line-of-sight merger.
If there is a line-of-sight merger, there are almost certainly
additional galaxies in the line-of-sight which artificially increase the
optical galaxy velocity dispersion if only one component is assumed.

\subsection{X-ray properties}

The mass estimates from the $M$--$T$ relation in Finoguenov et al. (2001)
and in Bryan \& Norman (1998) are
$M_{200} = 5.96 \times 10^{14}$~$h_{70}^{-1}{\rm M_{\odot}}$
and $M_{200} = 3.76 \times 10^{14}$~$h_{70}^{-1}{\rm M_{\odot}}$,
which indicate a higher gravitational mass.
We measure a bolometric X-ray luminosity of  
$2.9 \pm 0.1\times 10^{44} h^{-2}_{70}$~erg~s$^{-1}$, which is higher than 
the luminosity of $1.26\times 10^{44} h^{-2}_{70}$~erg~s$^{-1}$ 
obtained from the $L$--$T$ relation in Arnaud \& Evrard (1999)
which might indicate a merger.
The gas mass fraction at $r_{2500}$ for CL0024$+$17 agrees with 
the measurements of Allen et
al. (2002) and Pratt \& Arnaud (2002).
The XMM-Newton measured gas mass fraction of $0.20\pm0.03 h^{-3/2}_{70}$ 
at $r_{200}$ for CL0024$+$17 is higher than the values 
in Sanderson et al. (2003), Ettori et al. (2002) and Pratt \& Arnaud (2002) 
for their cluster samples, which is possibly an indication 
for an underestimated mass beyond $r_{2500}$ 
as already noted from the $M$--$T$ relation. Thus we suggest
that the merger leads to distortions on global scales which 
reduces the mass for CL0024$+$17. However, on global 
scales the effect is smaller than in the cluster center. 
The relatively small distortion (less than a factor of 2) was found 
in several simulations (e.g. Evrard et al. 1996; Schindler 1996).

\subsection{Central cooling}

CL0024$+$17 has a bright central galaxy and a compact lensing core.
The entropy in the center is a little below the entropy floor value
suggested by Lloyd-Davies et al. (2000).
However, the cluster almost shows little evidence for 
a cooling flow or a central temperature decrease 
which should be found in most cooling 
flow clusters. This supports the scenario that the cluster center has
been disturbed by a recent merger.

\section{Summary and Conclusions}
\label{s:conclusion}

We performed a detailed imaging and spectroscopic study 
of the XMM-Newton observations of the 
lensing cluster CL0024$+$17, and 
obtained spatially resolved temperature, metallicity  
and density distributions. A temperature gradient was observed for the first 
time in CL0024$+$17 at large radii ($1.3^{\prime}<r<3^{\prime}$). 

The image shows overlapping mass concentrations in the 
XMM-Newton X-ray data and HST optical data.
The substructure which was 
observed by HST in Kneib et al. (2003) is confirmed in the X-ray data.
The HRM shows an elongation in the NW-SE direction on a scale of 
$3.3^{\prime}$. Further spectroscopy confirms the temperature map
suggesting high temperature region with
$k_{\rm B}T=5.8\pm 1.0$~keV. Similar to Sersic~159$-$03, the 
observed temperature distribution shows a 
temperature gradient at $1.3^{\prime}<r<3^{\prime}$. 
The modeling of the temperature distribution yields a polytropic 
index of 1.17, lower than the adiabatic index of $\gamma=5/3$, which
indicates a convectively stable state in the outskirts of CL0024$+$17.

The cluster does not show a pronounced cooling flow 
in spite of a dominant central galaxy. Therefore we suggest 
that a merger in the cluster center ($<0.1h_{70}^{-1}$~Mpc) has disrupted any 
previous cooling flow.

The mass estimate based on the precise
measurements of the distributions of the temperature and gas density
is lower by a factor of $\sim $~20--25\% 
than the mass obtained under the assumption of
isothermality in the outskirts. The 
NFW model does not fit the X-ray derived mass profile very well
in the center ($r<0.1 h^{-1}_{70}$~Mpc). The extended NFW model 
fits the data but yields a negative index $\alpha$.
We attribute this to the disturbance of the core.

The discrepancy remains between the X-ray gravitational mass 
and optical strong lensing mass at the radii of $0.143h^{-1}_{70}$~Mpc 
(Broadhurst et al. 2000) and $0.153h^{-1}_{70}$~Mpc (Tyson et al. 1998)
by a factor of up to 4. 
The XMM-Newton results are marginally consistent
within $2 \sigma$ using
the global temperature with the weak lensing measurement 
in Kneib et al. (2003). CL0024$+$17 is an example of 
a cluster for which weak lensing masses 
and X-ray masses are in acceptable agreement at 
large radii but for which the strong lensing masses 
disagree with the X-ray derived values at small radii (also  
e.g. A1689, Xue \& Wu 2002; Dye et al. 2001; Clowe \& Schneider).

The XMM-Newton measured parameters for CL0024$+$17 
deviate slightly from the empirical
$M$--$T$ and $L$--$T$ relations. CL0024$+$17 has a bright central galaxy 
and a compact lensing core but does not show a 
significant cooling flow. Similar to A2218 
(Girardi et al. 1997; Pratt et al. 2004), 
the apparent discrepancy between the X-ray and gravitational
lensing determined mass of CL0024$+$17 is most probably due to 
a line-of-sight merger of two almost comparable subsystems. 
In addition, filamentary structures could also contribute to the 
projected mass detected by lensing, but since this mass does not 
lie within the cluster core, it would not be included in the X-ray 
mass measurement. 
The line-of-sight orientation makes it difficult to reveal the merger
structure in the X-ray data, but on the other hand it enhances 
the probability of finding strong lensing features. 
Thus, the application of 
hydrostatic equilibrium assumption might break down 
in the central region. On global scales the agreement becomes better. 
The galaxy distribution, lensing mass maps, and X-ray 
data all provide a consistent description of the cluster morphology.  

\begin{acknowledgements}
  
The XMM-Newton project is
supported by the Bundesministerium f\"ur Bildung und Forschung,
Deutschen Zentrum f\"ur Luft und Raumfahrt (BMBF/DLR), the Max-Planck
Society and the Haidenhaim-Stiftung. We acknowledge Jacqueline
Bergeron, PI of the XMM-Newton observation of the CDFS.
YYZ acknowledges receiving the International Max-Planck Research School
Fellowship. W. Forman acknowledges support from NASA Grant
NAG5-9942 and the Smithsonian Institution.
YYZ thanks Alexey Vikhlinin, Gabriel Pratt, Kyoko Matsushita,
Maxim Markevitch, Keith Arnaud, Xiang-Zhong Zheng, Yas Hashimoto, 
Takaya Ohashi and Tae Furusho for useful discussions.
\end{acknowledgements}

\bibliographystyle{aabib99}

\end{document}